# Searching for Islands of Reionization: A Potential Ionized Bubble Powered by a Spectroscopic Overdensity at $z = 8.7$

Rebecca L. Larson[1,12], Steven L. Finkelstein[1], Taylor A. Hutchison[2,3,12], Casey Papovich[2,3], Micaela Bagley[1], Mark Dickinson[4], Sofía Rojas-Ruiz[5,13], Harry C. Ferguson[6], Intae Jung[7,8], Mauro Giavalisco[9], Andrea Grazian[10], Laura Pentericci[10], and Sandro Tacchella[11]

[1] The University of Texas at Austin, Department of Astronomy, Austin, TX 78712, USA; rlarson@astro.as.utexas.edu
[2] Department of Physics and Astronomy, Texas A&M University, College Station, TX 77843-4242, USA
[3] George P. and Cynthia Woods Mitchell Institute for Fundamental Physics and Astronomy, Texas A&M University, College Station, TX 77843-4242, USA
[4] NSF's NOIRLab, 950 N. Cherry Ave., Tucson, AZ 85719, USA
[5] Max-Planck-Institut für Astronomie, Königstuhl 17, D-69117, Heidelberg, Germany
[6] Space Telescope Science Institute, Baltimore, MD 21218, USA
[7] Astrophysics Science Division, Goddard Space Flight Center, Greenbelt, MD 20771, USA
[8] Department of Physics, The Catholic University of America, Washington, DC 20064, USA
[9] University of Massachusetts, Amherst, MA 01003, USA
[10] INAF—Osservatorio Astronomico di Padova, Vicolo dell'Osservatorio 5, I-35122, Padova, Italy
[11] Department of Physics, Ulsan National Institute of Science and Technology (UNIST), Ulsan 44919, Republic of Korea
*Received 2021 November 10; revised 2022 February 18; accepted 2022 March 9; published 2022 May 9*

## Abstract

We present the results from a spectroscopic survey using the MOSFIRE near-infrared spectrograph on the 10 m Keck telescope to search for Lyα emission from candidate galaxies at $z \sim 9\text{--}10$ in four of the CANDELS fields (GOODS-N, EGS, UDS, and COSMOS). We observed 11 target galaxies, detecting Lyα from one object in ∼8.1 hr of integration, at $z = 8.665 \pm 0.001$ with an integrated signal-to-noise ratio > 7. This galaxy is in the CANDELS Extended Groth Strip (EGS) field and lies physically close (3.5 physical Mpc [pMpc]) to another confirmed galaxy in this field with Lyα detected at $z = 8.683$. The detection of Lyα suggests the existence of large (∼1 pMpc) ionized bubbles fairly early in the reionization process. We explore the ionizing output needed to create bubbles of this size at this epoch and find that such a bubble requires more than the ionizing power provided by the full expected population of galaxies (by integrating the UV luminosity function down to $M_{UV} = -13$). The Lyα we detect would be able to escape the predominantly neutral intergalactic medium at this epoch if our detected galaxy is inhabiting an overdensity, which would be consistent with the photometric overdensity previously identified in this region by Finkelstein et al. This implies that the CANDELS EGS field is hosting an overdensity at $z = 8.7$ that is powering one or more ionized bubbles, a hypothesis that will be imminently testable with forthcoming James Webb Space Telescope observations in this field.

*Unified Astronomy Thesaurus concepts:* High-redshift galaxies (734); Lyman-alpha galaxies (978); Reionization (1383); Galaxy spectroscopy (2171); Spectroscopy (1558); High-redshift galaxy clusters (2007)

## 1. Introduction

Observational studies of the reionization of the intergalactic medium (IGM) have predominantly focused on the end of this process. Quasar absorption spectra (e.g., Fan et al. 2006; McGreer et al. 2015) show that reionization was largely finished by $z \sim 6$ (perhaps some small neutral patches remain to $z \sim 5.5$; e.g., Becker et al. 2015, 2021). How fast reionization proceeds, what drives it, and how and when it began are poorly constrained. One scenario favors late reionization (e.g., Robertson et al. 2015), where UV emission from galaxies accounts for the ionizing photons needed to match the constraints. This model requires an ionizing photon escape fraction of 20% for all galaxies; if this is true, then $>0.01 L^{\star}$ galaxies dominate reionization (Naidu et al. 2020).

In contrast, observations find that the vast majority of galaxies have small (<5%) ionizing photon escape fractions (e.g., Siana et al. 2010; Grazian et al. 2017; Rutkowski et al. 2017) and galaxies with high escape fractions are rare (see Izotov et al. 2016). Simulations such as those from Paardekooper et al. (2015) predict that the escape fraction depends on halo mass, with higher escape from galaxies in the lowest-mass halos (log $M_h/M_\odot < 9$). In this second scenario, the faintest galaxies ($<0.01 L^{\star}$) dominate reionization, and the high abundance of faint galaxies at very early times (Mason et al. 2015; Finkelstein 2016) drives an earlier start to reionization and a smoother time evolution (Finkelstein et al. 2019). The most dramatic difference between these two scenarios is at $z \sim 8\text{--}9$. The late-reionization model, where *all galaxies contribute*, predicts ∼20% ionization of the IGM (by volume) at $z = 8\text{--}9$, while the early-reionization model, where *the extreme faint galaxies dominate*, predicts a ∼60% ionized fraction at these redshifts (e.g., Finkelstein et al. 2019). Late reionization should thus have only rare ionized patches at $z > 8$, compared to relatively common ionized patches for early reionization. As Lyα detectability is inversely proportional to the neutral fraction of the IGM, we would expect to see Lyα from galaxies at $z > 8$ more readily in the early-reionization model.

---

[12] NSF Graduate Fellow.
[13] Fellow of the International Max Planck Research School for Astronomy and Cosmic Physics at the University of Heidelberg (IMPRS–HD).

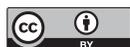

The evolution of this phase change process is largely unknown, but one of the best probes to use in constraining it is the spectroscopic measurement of Lyα emission from galaxies at this epoch. Due to its resonant scattering properties, Lyα has been a standard tool to probe the end of reionization at $z \sim 6$–7 (e.g., Miralda-Escudé & Rees 1998; Malhotra & Rhoads 2006; Dijkstra 2014). This method is possible because the fraction of Lyman-break-selected galaxies that have detected Lyα emission increases from ∼30% at $z = 3$ to 60%–80% at $z = 6$ (Shapley et al. 2003; Stark et al. 2010, 2011). It is expected that this trend in increased Lyα emission continues past $z = 6$, as early galaxies have lower metallicities and thus lower dust attention, allowing for bright UV emission to escape the galaxy. However, there is recent evidence that Lyα is not detected in as many galaxies at this epoch as is expected (Pentericci et al. 2011, 2014; Ono et al. 2012; Schenker et al. 2012; Tilvi et al. 2014). An observed drop in the fraction of these continuum-selected galaxies that emit Lyα may be indicative of a rapidly evolving IGM neutral fraction, as these Lyα photons are resonantly scattered by neutral gas and would not be detectable (see also Naidu et al. 2020).

However, there are notable exceptions, with many recent spectroscopic Lyα detections at $z = 7.0$–8.7, implying that ionized bubbles exist as early as $z > 8.5$ (Castellano et al. 2018; Mason et al. 2018; Jung et al. 2020; Tilvi et al. 2020; Wold et al. 2022). Many of the $z > 7.5$ galaxies with Lyα detections exhibit indications of strong [O III] emission (inferred from their red Spitzer IRAC [3.6]−[4.5] colors; e.g., Oesch et al. 2015; Zitrin et al. 2015; Roberts-Borsani et al. 2016; Stark 2016), indicating high ionization parameters. If these photons can leak from these relatively massive galaxies because of their high ionization rates, this could be a bias in the detectability of Lyα, as they may be driving the formation of ionized bubbles in the IGM. Or, because these galaxies also likely trace overdensities (e.g., Castellano et al. 2016; Endsley et al. 2021), it may be their fainter neighboring galaxies that leak UV light and ionize the bubbles. In either scenario, the massive galaxies sit in ionized bubbles large enough for Lyα to redshift out of resonance by the time it reaches the edge, which implies a bubble size of at least 1 physical Mpc (Malhotra & Rhoads 2004; Dijkstra 2014).

However, detection of some Lyα emission from smaller ionized regions may be possible if the bulk of the Lyα photons exhibit a high-velocity offset ($\Delta v_{Ly\alpha} \gtrsim 300$ km s$^{-1}$) from the systemic redshift of the galaxy (Mason & Gronke 2020). There is thus a degeneracy between any derived bubble size, the assumed escape fraction of ionizing photons, and the velocity offset. While future measurements of line profile shapes in this epoch will allow more complex studies of bubble morphologies, even with current facilities we may be capable of inferring the presence of ionized regions by the sheer detection of Lyα.

Further evidence comes from observations of the Lyα luminosity function (LF). While some previous work has shown that the Lyα LF declines from $z = 5.7$ to $z \sim 7$ (e.g., Konno et al. 2014; Inoue et al. 2018), the bright end of the $z \sim 7$ Lyα LF does not exhibit this decline (Zheng et al. 2017; Wold et al. 2022). This is consistent with the bright $z \sim 7$ galaxies inhabiting ionized bubbles, leading to high Lyα detection rates (e.g., Mason et al. 2018; Jung et al. 2020; Tilvi et al. 2020; Endsley et al. 2021). Additionally, follow-up studies of $z \gtrsim 7$ Lyα detections have found evidence that multiple bright galaxies inhabit the same bubbles, with multiple galaxies exhibiting detectable Lyα at $z > 7$ (Castellano et al. 2018; Tilvi et al. 2020). If the early-reionization model from Finkelstein et al. (2019) is correct, we would expect to see the same situation at a higher redshift, with bright $z > 8$ galaxies occupying large ionized regions, such that we could detect Lyα emission from them. However, if the late-reionization model from Robertson et al. (2015) is correct, most $z > 8$ galaxies would still be surrounded by a neutral IGM and thus not detectable in Lyα.

The suggestion of the existence of ionized bubbles at $z > 7$ begs the question—are there large ionized bubbles at higher redshift? We exploit a new sample of 11 bright, photometrically selected galaxies at $z \sim 9$–10 to perform deep follow-up spectroscopy to search for Lyα from this distant epoch. Although challenging due to the faint nature of these sources and the ubiquitous telluric emission lines in ground-based data at these wavelengths, the detection of even a single galaxy in Lyα would indicate the presence of an ionized bubble deep in the epoch of reionization. While a single line (or bubble) cannot easily distinguish between the two model scenarios, showing that Lyα is detectable in this epoch will further motivate wide-area Lyα searches with future facilities such as the Nancy Grace Roman Space Telescope and the Giant Magellan Telescope.

This paper is organized as follows: In Section 2 we describe our observations and the Islands of Reionization survey. Section 3 follows with a step-by-step description of our unique data reduction process, including improvements made to the pipeline and details on the one-dimensional (1D) spectra extraction and flux calibration processes. In Section 4 we walk through our data analysis and emission-line search method, including criteria required for successful detections. Section 5 is where we present our results and the detection of Lyα in a galaxy at $z = 8.665$, and we discuss the implication of our detection in Section 6. Concluding remarks are found in Section 7. Where relevant we assume the cosmological parameters from Planck Collaboration et al. (2020), using their TT, TE, EE + lowE + lensing model (Table 2, Column (5)). For $H_0$ we use 67.36, $\Omega_b = 0.02237/h^2$, $\Omega_\Lambda = 0.6847$, $\Omega_m = 0.3153$, and the helium fraction $Y_p = 0.2454$.

## 2. Observations

We obtained Keck/MOSFIRE $J$-band spectroscopic follow-up targeting Lyα from a sample of candidate $z = 9$–10 galaxies recently published by Finkelstein et al. (2022), discovered in the Cosmic Assembly Near-infrared Deep Extragalactic Legacy Survey (CANDELS; Grogin et al. 2011; Koekemoer et al. 2011). The full details of their selection are discussed in Finkelstein et al. (2022), but, in brief, these galaxies were selected as those with significant ($>7\sigma$) photometric detections, with photometric redshifts that were strongly constrained to be at $z > 8$. Finkelstein et al. (2022) detailed several vetting procedures done to ensure the robustness of the sample, culminating in a sample of 14 initial $z = 9$–10 candidates, which were honed to 11 robust candidates following the inclusion of ground-based imaging and additional Hubble Space Telescope (HST) F098M and F140W imaging to rule out low-redshift solutions. Figure 1 shows the apparent magnitude distribution of the final sample of galaxies in this survey. We note that while the search included all five CANDELS fields, GOODS-S held no $z > 8$ galaxies that met the robust detection criteria. We thus continue our survey over the remaining four CANDELS fields (GOODS-N, EGS, UDS, and COSMOS).

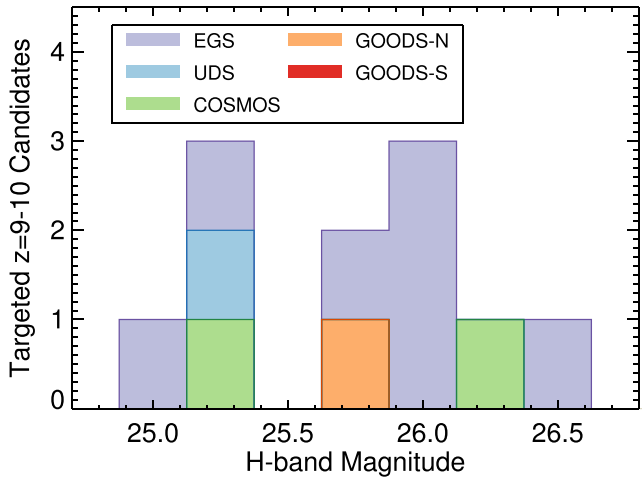

**Figure 1.** *H*-band (F160W) apparent magnitude and field distribution of the final sample of $z = 9–11$ galaxies from Finkelstein et al. (2022) across the five CANDELS fields. No such candidates were identified in the GOODS-South field, and thus none of our MOSFIRE observations targeted this field. Note: 7 of our 11 targets are located in the EGS field.

Our spectroscopic data were obtained over a total of 16 separate calendar nights of observations over four CANDELS fields (GOODS-N, EGS, UDS, and COSMOS) with Keck/MOSFIRE through the Islands of Reionization program (PI: R. Larson; awarded as eight nights over four semesters through the NASA/Keck allocation). We obtained additional data for one target in the CANDELS/GOODS-N field through a program led by T. Hutchison, awarded as two nights through the NASA/Keck allocation in 2020 February. An additional hour of observation was taken in the EGS field during a program led by PI J. Bridge in 2018 May. Details for each night of observations can be found in Table 1.

Here we analyze the entire MOSFIRE data set with 18 mask designs from 2018 April to 2020 February. The positions of our masks and slits are shown in Figure 2. To detect Ly$\alpha$ over a redshift range of $8.5 < z < 11$, we used the *J*-band filter, which covers $\lambda = 1.153–1.352\,\mu\text{m}$ at a spectral resolution of $R \approx 3300$. The slit width was chosen to be $0''\!.7$ to match the typical seeing level on Maunakea. During our observations, individual frames were taken with 120 s exposures with an ABAB dither pattern ($+1''\!.25, -1''\!.25, +1''\!.25, -1''\!.25$). The details of the observations are described in Table 1. We observed 9 of the 11 galaxies from the final sample of Finkelstein et al. (2022) with this program (which also includes several sources previously published by Oesch et al. 2018; Bouwens et al. 2019). Of the two sources not observed, we explicitly avoided EGS_z910_6811,[14] as this is EGSY8p7 from Zitrin et al. (2015) and already has a spectroscopic redshift from Ly$\alpha$ at $z = 8.683$ from MOSFIRE, while source EGS_z910_68560 lies far from the other targeted sources and would have required its own pointing. We list the observed galaxies in Table 2, including the three candidates that Finkelstein et al. (2022) removed from their final sample owing to additional imaging obtained that pushed the photometric redshift calculations to lower redshifts, but which we still observed here (see the Appendix).

---

[14] Here we use the identifications from Finkelstein et al. (2022), which include both the field name and their catalog ID number.

## 3. Data Reduction

Through rigorous testing and after noticing a drift of our objects in their slits throughout a long night of observations, we found it necessary to modify our use of the Keck/MOSFIRE data reduction pipeline (DRP)[15] to reduce our data. Below we discuss the changes we made in our data reduction processes and attempt to provide a resource for others who might want to use this instrument to search for similarly faint emission lines. We follow similar steps to those in Jung et al. (2020), where we track the drift of objects in the slit through the night (Section 3.2) and correct for it (Section 3.2.1). We also implement some additional steps owing to the faint nature of our sources and expected detections, such as initial removal of cosmic rays (CRs; Section 3.1), a triple-Gaussian extraction profile (Section 3.3), a rolling extraction at different centers (Section 3.3.1), and masking of slit contamination (Section 3.3.2).

### 3.1. Cosmic-ray Removal

The DRP requires at least five frames to effectively remove CRs. However, as we discuss below, we run the DRP on pairs of frames, which means that we must clean CRs or bad pixels separately. To do this, we first run each raw frame through the Python version of LA Cosmic (van Dokkum 2001), with the following parameters set: gain = 2.15, readnoise = 10, sigclip = 9.7, sigfrac = 0.14, objlim = 5.85, satlevel = 50,000. We also employ sigma-clipping rejection when we combine reduced frames for each night to account for any missed CRs, specifically because LA Cosmic is less effective at identifying large CRs.

### 3.2. Accounting for Slit Drift

Previous MOSFIRE observations have reported a noticeable drift along the spatial direction of the slit (e.g., Kriek et al. 2015; Song et al. 2016; Jung et al. 2019, 2020; Hutchison et al. 2020) throughout a night of observation. We detect similar drifting in our data, which must be corrected, as we want to prevent the smearing out of our source signal and ensure accurate and robust detections. Hutchison et al. (2020) determined that this drift is due to uncorrected residuals from the instrument's flexure compensation systems. In particular, a model is used to correct flexure between the guide camera and science field, but there appear to be hysteresis effects that are uncorrected by the model, and those dominate the pixel-scale drifts seen in the data. To correct for this slit drift, we reduced each adjacent pair of science frames separately, generating reduced two-dimensional (2D) spectra for each 240 s of integration time. We reduced the raw data using the 2018 version of the public MOSFIRE data reduction pipeline. The DRP provides a sky-subtracted, flat-fielded, and rectified slit spectrum per object with a wavelength solution based on telluric sky emission lines. In the reduced 2D spectra, the spectral dispersion is 1.09 Å pixel$^{-1}$, and the spatial resolution is $0''\!.18$ pixel$^{-1}$. We had placed stars in slits to assist in flux calibration, but we were also able to use them as continuum sources with which we could track the drift in our observations. In the instances where we had two (or more) stars in slits for a mask, we chose the star with the least variability and drift throughout the night. For masks that were observed on multiple nights, we used the same

---

[15] http://keck-datareductionpipelines.github.io/MosfireDRP/

Table 1
MOSFIRE Masks Observed

| Mask # | Mask | Date | R.A. (J2000) | Decl. (J2000) | PA (deg) | Exp. Time (hr) | Avg. Seeing (arcsec) |
|---|---|---|---|---|---|---|---|
| E1 | EGS_2018A_J_Mask1 | 2018 Apr 25 | 14:20:45.05 | +53:01:38.49 | 52.00 | 4.90 | 0.64 |
| (E2) | EGS_2018A_J_Mask1_2 | 2018 May 21 | 14:20:45.05 | +53:01:38.49 | 52.00 | 1.33 | 1.11 |
| E3 | EGS_2019A_J_1 | 2019 Mar 26 | 14:20:43.11 | +53:02:59.51 | 93.50 | 3.18 | 0.69 |
| E4 | EGS_2019A_J_2 | 2019 Mar 27 | 14:19:40.33 | +52:53:29.71 | 48.00 | 3.25 | 0.96 |
| E5 | EGS_2019A_J_2-2 | 2019 Mar 28 | 14:19:40.34 | +52:53:29.61 | 48.00 | 3.11 | 1.10 |
| E6 | EGS_2019A_J_2-3 | 2019 Mar 29 | 14:19:40.33 | +52:53:29.71 | 48.00 | 3.18 | 1.25 |
| (G1) | GOODSN_J_Mask1 | 2018 Apr 25 | 12:37:12.71 | +62:18:40.88 | −39.70 | 0.99 | 0.70 |
| G2 | GOODSN_2020A_J | 2020 Feb 27 | 12:36:46.59 | +62:16:11.89 | 274.00 | 3.71 | 0.97 |
| U1 | UDS_MOSFIRE_2018B_J | 2018 Nov 25 | 02:17:14.83 | −5:09:50.74 | 115.90 | 5.90 | 0.95 |
| U2 | UDS_MOSFIRE_2018B_2 | 2018 Dec 12 | 02:17:14.83 | −5:09:49.44 | 115.70 | 4.51 | 1.12 |
| U3 | UDS_2019B | 2019 Dec 12 | 02:17:21.63 | −5:14:56.77 | 122.80 | 3.84 | 0.68 |
| U4 | UDS_2019B_2 | 2019 Dec 13 | 02:17:21.63 | −5:14:56.77 | 122.80 | 3.98 | 0.94 |
| C1 | COSMOS_MOSFIRE_2018B_J | 2018 Nov 25 | 10:00:26.03 | +2:13:49.06 | 180.90 | 2.12 | 0.70 |
| C2 | COSMOS_MOSFIRE_2018B_2 | 2018 Dec 12 | 10:00:25.82 | +2:13:51.26 | 179.40 | 3.25 | 0.83 |
| C3 | COSMOS_2019B | 2019 Dec 12 | 10:00:28.91 | +2:24:52.40 | 230.00 | 2.98 | 0.64 |
| (C4) | COSMOS_2019B_2 | 2019 Dec 13 | 10:00:28.91 | +2:24:52.40 | 230.00 | 0.27 | 0.88 |

**Note.** List of masks observed. Most were through our Islands of Reionization program, except for the 2018 May 21 observation in EGS (E2), which was observed during a separate program (PI Bridge), and the GOODS-N observation in 2020 February (G2), which was part of a program searching for nebular UV metal lines at high redshift (PI: T. Hutchison). Total exposure time and average seeing over the time of integration are indicated for each mask. Coordinates indicate the central coordinates and the position angle (PA) for the mask as input into the MOSFIRE observation software. Masks that have exposure times of less than 2 hr (marked in parentheses) were excluded from our analysis, as their inclusion increased the noise in our final stacks (note: including these data did not change our final detection results). Masks that do not contain the final sample of Finkelstein et al. (2022) candidate galaxies but were observed are listed in Table 8 in the Appendix.

slit star each time. A list of the stars we placed on slits, and their measured properties can be found in Table 4.

We started with the first set of frames and derived 1D cross dispersion profiles using the stars we observed by collapsing the reduced 2D spectra of the stars (from the MOSFIRE DRP) in the spectral direction and using the median value to produce a 1D spatial profile of each star. We fit a Gaussian to the slit star's spatial profile and used the results of these fits for our measurements of each pair of frames for the night. Offsets in the spatial direction from the first frame are taken from the pixel location of the peak of the Gaussian fit. The seeing for each pair of frames is estimated as the FWHM of the Gaussian fit to the slit star's profile, and the peak of the Gaussian fit is used as the weight for that pair of frames when creating the nightly stacks. This process is repeated for each set of observational frames for each mask for each night. Figure 3 shows the measured offsets in the spatial direction from the first frame as a function of time, showing drifts up to $\sim 0\rlap{.}''1\ \mathrm{hr}^{-1}$. As the slit drift in the cross dispersion direction was a known issue in MOSFIRE observations (Kriek et al. 2015; Jung et al. 2020), we tracked the drifting in real time during our observations and aligned the telescope when the drift became major; this is seen as the large jumps in the figure at ∼4.5 hr.

### 3.2.1. Shifting and Stacking Each Night

Once we measured the offset of each pair of reduced frames, the next step was to shift all the frames by this offset and combine the total frames for the mask for each night. To achieve an optimal signal-to-noise ratio (S/N), we combine each weighted DRP output frame (which is 240 s of combined exposure time) by a quantity related to the observing conditions. We tested three weighting methods: weighting by the peak of a Gaussian fit to the continuum-detected slit stars, the integral of this Gaussian, and the FWHM of this Gaussian. Additionally, we tested each of these both with and without removing frames with measured seeing worse than $1\rlap{.}''0$. We found that weighting by the peak of this Gaussian fit and NOT clipping out bad frames (seeing $> 1\rlap{.}''0$) yielded the lowest background noise in our final stacked spectra, and we thus used this method on our data. This follows the method laid out in Kriek et al. (2015). An example of the relevant properties measured from the data for one mask during one night is shown in Figure 4.

### 3.3. Extracting 1D Spectra for Nightly Stacks

The 1D spectra of the sources were obtained via an optimal extraction (Horne 1986) with a $1\rlap{.}''4$ spatial aperture, twice the typical seeing FWHM level from our observations. For the optimal extraction, we build a spatial weight profile from the slit star's trace so that the pixels near the peak of the trace are maximally weighted. Our dither pattern creates a triple-Gaussian profile for all real detections: a positive trace centered on the object, and two negative traces due to the positive flux in the frames used for background subtraction, each at n pixels to either side depending on the size of the dither. We elected to use this entire profile, both the positive and the negative, as our extraction profile (see Figure 5, left panel). This should enhance the S/N for real sources (which would be present in all exposures) versus spurious sources such as a noise peak, which would be present in just a single exposure. We note that this will artificially increase the flux in our spectra, but we correct for this in our flux calibration by scaling our flux calibration array to match measured CANDELS magnitudes for our slit stars (see Section 3.4). A discussion about how this extraction profile ultimately does not impact our measured line flux can be found in Section 5.2.

We dithered at $1\rlap{.}''25$, so our negative traces are located 14 pixels above and below the positive trace in the slit (in the 2018 November observations [Masks U1 and C1] we accidentally set our dither at $1\rlap{.}''5$, and thus the trace separation is 16 pixels). We

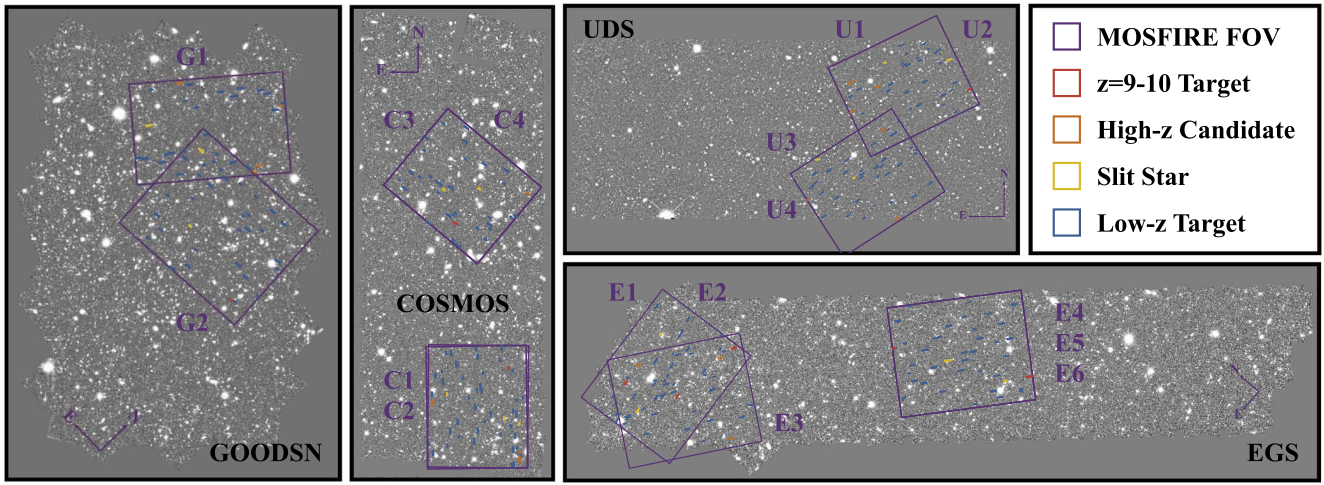

**Figure 2.** Locations of our MOSFIRE observations over four of the CANDELS fields are shown with the field of view of MOSFIRE as purple squares and labeled by mask numbers from Table 1. Final $z = 9$–10 targets from Finkelstein et al. (2022) are marked in red; early candidates in our target sample that were later removed are marked by orange slits (see the Appendix). The stars we put slits on for flux calibration and tracing the drift of our objects are marked in yellow. Low-redshift targets used to fill remaining slits are marked in blue. When masks overlap, it indicates multiple calendar nights of observations that we stacked as detailed in Section 3.5.

**Table 2**
Observed $z = 9$–10 Target Galaxies

| Galaxy ID | R.A. (J2000) | Decl. (J2000) | $H_{F160W}$ (mag) | $z_p$ | Masks Observed | Total Exp. Time (hr) | Avg. Seeing (arcsec) |
|---|---|---|---|---|---|---|---|
| EGS_z910_44164 | 215.218737 | 53.069859 | 25.41 ± 0.07 | $8.87^{+0.19}_{-0.17}$ | E1, (E2), E3 | 8.08 (9.41) | 0.67 (0.81) |
| EGS_z910_26816 | 215.097775 | 53.025095 | 26.11 ± 0.10 | $9.38^{+0.28}_{-0.39}$ | E1, (E2), E3 | 8.08 (9.41) | 0.67 (0.81) |
| EGS_z910_20381 | 215.188415 | 53.033644 | 26.05 ± 0.10 | $8.67^{+0.32}_{-0.74}$ | E3 | 3.18 | 0.69 |
| EGS_z910_26890 | 214.967536 | 52.932966 | 26.09 ± 0.07 | $8.99^{+0.22}_{-0.29}$ | E4, E5, E6 | 9.54 | 1.10 |
| EGS_z910_40898 | 214.882993 | 52.840414 | 26.50 ± 0.11 | $8.77^{+0.25}_{-0.90}$ | E4, E5, E6 | 9.54 | 1.10 |
| GOODSN_z910_35589 | 189.106061 | 62.242040 | 25.82 ± 0.05 | $10.41^{+0.24}_{-0.08}$ | G2 | 3.71 | 0.97 |
| UDS_z910_18697 | 34.255636 | −5.166606 | 25.32 ± 0.09 | $9.89^{+0.16}_{-0.15}$ | U1, U2 | 10.41 | 1.04 |
| COSMOS_z910_20646 | 150.081846 | 2.262751 | 25.42 ± 0.08 | $9.80^{+0.10}_{-0.46}$ | C1, C2 | 5.37 | 0.77 |
| COSMOS_z910_47074 | 150.126386 | 2.383777 | 26.32 ± 0.10 | $9.64^{+0.22}_{-0.13}$ | C3, (C4) | 2.98 (3.25) | 0.64 (0.76) |
| UDS_z910_731* | 34.317089 | −5.275935 | 25.02 ± 0.12 | $3.72^{+0.03}_{-0.03}$ | U3, U4 | 7.82 | 0.81 |
| UDS_z910_7815* | 34.392823 | −5.259911 | 25.82 ± 0.15 | $9.66^{+0.27}_{-7.29}$ | U3, U4 | 7.82 | 0.81 |
| COSMOS_z910_14822* | 150.145769 | 2.233625 | 24.51 ± 0.04 | $2.16^{+0.06}_{-0.25}$ | C1, C2 | 5.37 | 0.77 |

**Note.** The list of target $z = 9$–10 galaxies observed during our MOSFIRE program from the sample in Table 3 of Finkelstein et al. (2022). Two galaxies in the final sample were not observed in our MOSFIRE program, one because it was in an isolated region of the field (EGS_z910_68560), and the other (EGS_z910_6811) because it already contained a robust Ly$\alpha$ detection in Zitrin et al. (2015). The asterisk indicates galaxies that had $z > 8$ redshifts preferred with the initial HST+IRAC photometry but were removed from the Finkelstein et al. (2022) sample after ground-based photometry and additional HST photometry were added to their photometric redshift calculations. Mask names in which each galaxy was observed (from Table 1), as well as total exposure time and average seeing over all observed nights, are shown for each galaxy. For those masks that are not included in our sample owing to exposure times less than 2 hr, we include them in parentheses for completeness, but they are not included in our final analysis.

used each mask's chosen slit star (see Section 3.2) to create the extraction profile, made by collapsing the star's 2D S/N spectrum in the spectral direction, taking the median value at each spectral pixel, and fitting a triple Gaussian to the positive trace and two negative traces. We found that the negative traces were not always equal to each other in peak flux, likely due to situations where the observations were stopped, leaving an uneven number of dither A and B frames. Thus, we measured the flux of the positive trace in the slit and then used $-\sqrt{2}$ times this value as the flux of the negative traces for our profile. This ensures that we have a total weight of 1 when integrating over the whole extraction profile. We also note that for the stars that do have equal negative traces the negative Gaussian integrals are equal to $\sim -\sqrt{2} \times$ the positive Gaussian; thus, our corrections are in line with the flux of the real object. We then line up the center (peak of positive trace) with the row we want to use as the center for the object's extraction, adding or subtracting blank rows to make the model the same size as the slit. This is then made into an array the same length as the 2D spectrum, which is always 1528 pixels for the MOSFIRE J-band filter (McLean et al. 2012). Our 2D spectra are then extracted into 1D spectra using this profile as the weight. We note that we are assuming that our targets are point sources like the slit stars we are using as the extraction profile, and at this high redshift this is a reasonable approximation for our galaxies, as they are small, faint, and unresolved.

*3.3.1. Rolling Extraction*

When correcting for the shift of our slit stars throughout each night, we noticed that our continuum-detected objects were offset spatially from where the MOSFIRE DRP expected them to be. We tested many different masks, using the slit stars and all the low-redshift filler targets that had continuum detections, and could find no discernible pattern for the offsets, as they were all offset from the DRP expected location in different directions and amounts within the same mask. We observed that none of the objects were centered more than 3 pixels from the expected position, including the stars we have on slits. The distribution of offsets of our slit stars is shown in Figure 6, where we show all the stars, as often we had two per mask, and the offsets of the slit stars we chose to use for our data reduction.

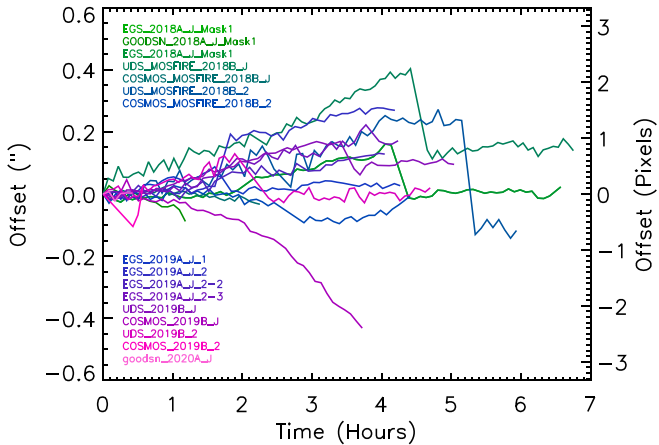

**Figure 3.** A known complication with MOSFIRE observations is that the objects drift in their slits spatially throughout long exposure times. Often this can be corrected by realigning the telescope, but this is time-consuming and does not fully correct for the drift. We measured offsets in the spatial direction of the peak of our slit stars from the first frame as a function of time for each mask of observations and plot them as both pixel and arcsecond offsets. Different colors represent different masks and are listed in chronological order. Large jumps in offset correspond to when we realigned the telescope. We shift each of our observation frames to correct for this drift during our reduction process. Details about this can be found in Section 3.2.

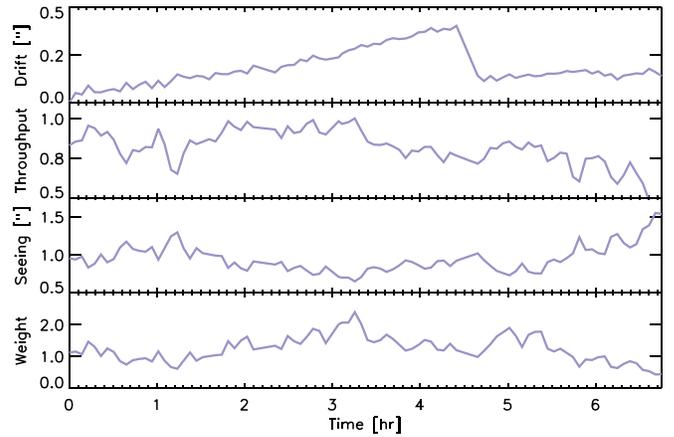

**Figure 4.** Measured characteristics of the data from Mask U1 taken on 2018 November 25. This mask is shown because it covers the longest integration time in a single night; the purple represents the star chosen for our measurements (s17875; see Table 3). For each of our pairs of frames we fit a Gaussian to the slit star's spatial profile and use the results of these fits for our measurements. From top to bottom: measured drift in the spatial direction from the first frame as a function of time, which is the offset of the peak of the Gaussian fit; throughput which is the flux relative to the maximum flux measured for that star for that mask; seeing which is the FWHM of the Gaussian; the weight used when stacking, which is the peak of a Gaussian fit (see Section 3.2). This plot is similar to that of Figure 3 in Kriek et al. (2015), where they also modified the MOSFIRE DRP to account for the drift of their objects through the night.

For objects that are easily distinguishable in one night's observations, finding the trace and centering an extraction profile would be simple. However, our $z = 9$–10 galaxies require multiple nights of observations to detect the expected emission strength of Ly$\alpha$. To combine multiple nights of observations and increase our exposure time to that required for detection, we must first locate our galaxy in the slit and extract to 1D, flux-calibrate, and then combine with the other nights. However, knowing that our object is potentially not where it is expected to be in the slit, and not being able to determine the exact offset but only a range of potential offsets, we must then extract from multiple possible positions in the slit and combine the individual nights with a variety of extraction centers, to account for all plausible locations of the galaxy spectrum in our slit. Based on the variation in slit positions from our slit stars (Figure 6), we used 7 different centers (Figure 5, left) in the slit for our extraction profile, 0 being the DRP expected center and then ±3 pixels above and below that center. This provided seven spectra per source per night of observation, centered on different potential locations of our objects in the slit. We explain in Section 3.5 below how this combination was performed.

*3.3.2. Dealing with Contamination in the Slit*

Occasionally objects other than our target fell within our MOSFIRE slits. We were careful to orient our masks such that the positive trace of potential contaminants would not interfere with the trace of our desired target, but because we are extending our extraction profile to include the negative traces from the dithers, we encountered a complication when the negative trace of the contaminating object overlapped with the negative trace of our object. This is the case with our target z910_44164 in the EGS field, and to account for this we removed one of the negative traces from our extraction profile. Since only one of the negative traces was contaminated in this object we are able to use the ratio of the integral of the new, contamination-removed extraction profile to that of the full, triple-Gaussian extraction profile as a scaling factor to apply to the resulting flux of our object to account for the "missing" negative trace. This ensures that we are accurately extracting the flux of our object without artificially increasing it by excluding the negative trace of the contaminating object.

### 3.4. Flux Calibration

We performed long-slit observations of spectro-photometric standard stars for flux calibration and telluric absorption correction using Kurucz (1993) model stellar spectra. We provide significant detail on our flux calibration process here in an effort to benefit readers who would like to follow our process. We chose standard stars from the list recommended by the MOSFIRE documentation,[16] preferentially selecting those which would be observed at nearly the same airmass where we would be observing our targets. We used the LONGSLIT setup with a $3'' \times 0''\!.7$ slit size and did 8 repeats of 12 s exposures using the ABA'B' dither pattern with $2''\!.25$ and $4''\!.0$ dithers. With the MOSFIRE DRP we reduce our standard star observations into 2D, sky-subtracted, wavelength-calibrated spectra. We then do a standard Gaussian extraction profile to extract a 1D spectrum of the standard star. These spectra are in units of [counts s$^{-1}$], so we then multiply them by the total exposure time taken from finding the maximum value in the "_itime.fits" file created by the pipeline.

---
[16] https://www2.keck.hawaii.edu/inst/mosfire/exposure_recipes.html

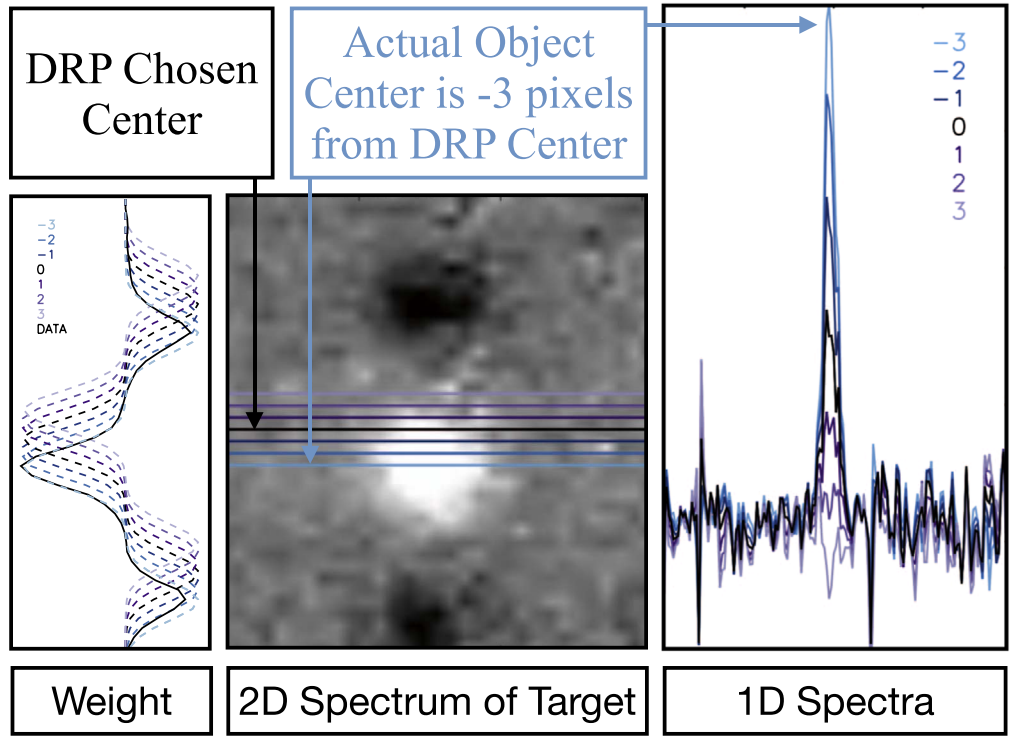

**Figure 5.** Different centers for the extraction profiles and the impact on the final extracted 1D spectrum. Often the MOSFIRE DRP suggested spatial row for objects was not the actual position of the object in the slit, so we performed a rolling extraction, centering our weight profile at seven different locations (left). For bright, low-redshift emission lines such as the one shown here (middle) we can determine the correct location to center our extraction by eye as where we get the highest flux in the emission line in the 1D spectrum (right, −3 pixels from the DRP center in this case). In the case of high-redshift targets, they are too faint to be detected in one night in order to identify the correct center, so we must use all possible extraction centers for stacking purposes. The correct extraction center for our weight profile ensures that we are getting the full flux of our emission lines and the highest S/N.

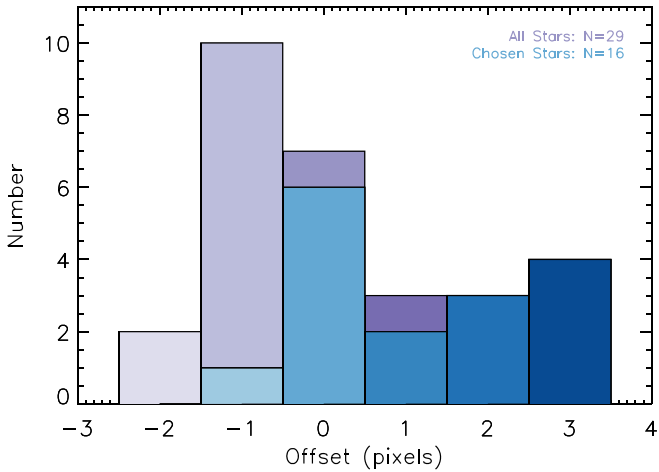

**Figure 6.** The measured offset in the spatial locations of our slit stars from their observed position to where the MOSFIRE DRP expected them to be. There was no discernible pattern to the offsets of our sources from the expected position (with, e.g., slit position, rotation angle, etc.), so we used the measured offsets of our slit stars as a guide. In purple we show a histogram of all slit-star offsets, and in blue we show the offsets of the slit stars we used for each mask (one per mask) in our data reduction and analysis. None of our slit stars were offset more or less than 3 pixels to either side of the expected location in the slit; thus, we center our extraction profile for our target galaxies at seven different centers: the DRP center pixel and ±3 pixels from that position (see Figure 5).

We obtained the spectral type and magnitude in the *J*-band for each standard star from the 2MASS catalog (Skrutskie et al. 2006), and convert them from Vega to AB magnitudes by adding 0.91 mag. We then read in the appropriate Kurucz model (Kurucz 1993) for the spectral type determined by 2MASS and convert this from $f_\lambda$ to $f_\nu$ units and interpolate this model and the 2MASS filter curves onto the standard star's wavelength array (all in Å). Next we integrate the Kurucz model through the 2MASS filter and calculate the appropriate scaling for our standard star, $\frac{f_{\rm 2MASS}}{f_{\rm model}}$, and multiply our model spectrum by this factor to match the known 2MASS *J*-band AB magnitude. Since there are some absorption features in our stellar spectra in the *J*-band (namely the absorption at ~1.28 μm) we interpolate over this absorption feature in both the model (orange, Figure 7 Left, Top) and the standard star spectrum. We further smoothed our standard star spectra with a width of 40 pixels to remove noise and any remaining stellar features (orange, Figure 7 Left, Middle). To obtain the response curve as a function of wavelength, we divided the interpolated model spectrum by the smoothed, interpolated standard star spectrum (orange, Figure 7 Left, Bottom).

In spite of our effort to match airmasses, our science masks were sometimes observed in different observing conditions than the long-slit standard stars due to changing atmospheric conditions. To mitigate this, we derived a scale factor to apply a residual normalization correction to make our calibrated spectra have the same *J*-band magnitude as that measured for our slit stars by CANDELS. This also corrects for slit loss, the effects of potential differences in ground- versus space-based fluxes, and for the boost in flux from our triple-peaked

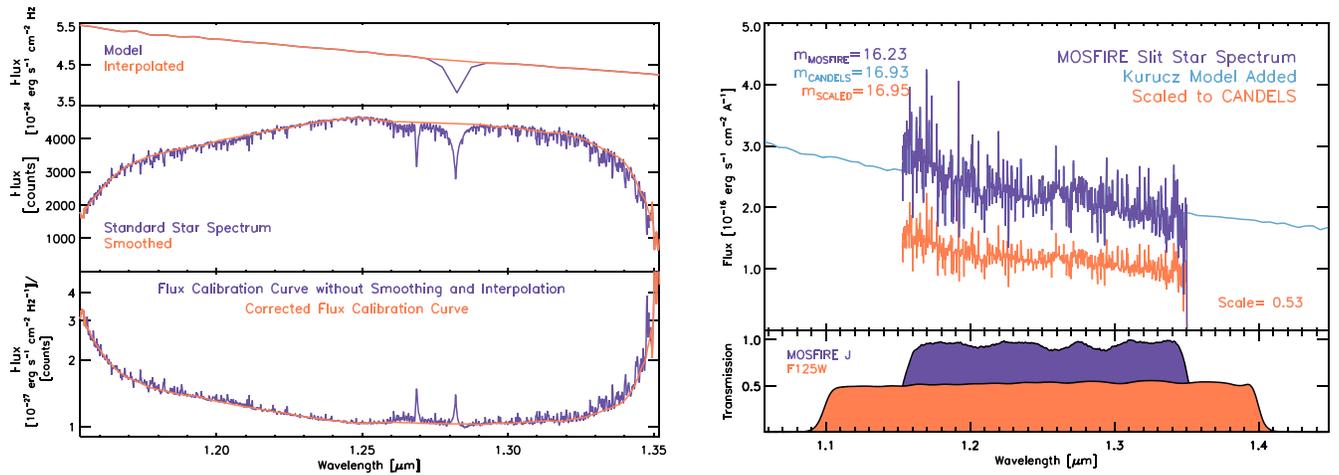

**Figure 7.** Left: standard star spectrum (middle, purple) and Kurucz model (top, purple) for HIP56736 from Mask U3. Both spectra are smoothed and interpolated (orange) to remove any absorption features and remove noise. The flux calibration array (bottom) is shown both with (orange) and without this step (purple). Final flux calibration array used is shown in orange. Right: spectrum of a slit star from our mask (purple, here is s14153 from mask U3). Since the MOSFIRE J filter covers a narrower range of wavelength than the CANDELS $J_{F125W}$ we use the Kurucz model to extend the spectrum and pass it through the $J_{F125W}$ filter. The magnitude measured is shown in purple and the known CANDELS magnitude is shown in blue. We then scale our slit star to match the magnitudes and pass it back through the $J_{F125W}$ filter as a check and show the scaled magnitude in orange. Description of this process can be found in Section 3.4. Standard stars used in our flux calibration are listed in Table 4 and slit stars and scale factors applied to our flux calibration arrays are listed in Table 3.

**Table 3**
Slit Stars Used for Scaling Flux Calibration and Tracing Drift of Targets Each Night

| Field | Star Catalog ID | Masks Observed | R.A. (J2000) | Decl. (J2000) | Total $J_{F125W}$ Magnitude | Type | Best Fit Kurucz Model | Scale Applied Per Mask |
|---|---|---|---|---|---|---|---|---|
| EGS | 156512 | E1, E2, E3 | 215.236280 | 53.049301 | 18.71 | G8V | kp00_5500[g45] | 0.22, 0.82, 0.44 |
| EGS | 318829 | E4, E5, E6 | 214.908890 | 52.853233 | 16.65 | F5V | kp00_6500[g40] | 0.32, 0.89, 1.21 |
| GOODS-N | 23900 | G1 | 189.367250 | 62.283123 | 19.89 | F5V | kp00_6500[g40] | 0.34 |
| GOODS-N | 31539 | G2 | 189.286990 | 62.296989 | 17.12 | G8V | kp00_5500[g45] | 0.41 |
| UDS | 17875 | U1, U2 | 34.273811 | −5.142240 | 16.95 | G8V | kp00_5500[g45] | 0.34, 1.05 |
| UDS | 14153 | U3, U4 | 34.353947 | −5.241657 | 16.93 | K0V | kp00_5250[g45] | 0.53, 0.24 |
| COSMOS | 2551 | C1, C2 | 150.085740 | 2.220684 | 17.24 | G2V | kp00_5250[g45] | 0.19, 0.32 |
| COSMOS | 14093 | C3, C4 | 150.106460 | 2.412378 | 20.07 | G8V | kp00_7000[g40] | 0.18, 0.70 |

**Note.** ID numbers come from CANDELS catalog numbers from Finkelstein et al. (2022). A flux-calibrated spectrum of each slit star was fit using the Kurucz (1993) models to determine the surface temperature and log(*g*) which were then used to determine the star's spectral types. These stars are those that were used in tracing the drift through the night and the profile is used to determine the weighting for each set of frames combined for each night of observations. Scales are shown to match MOSFIRE magnitudes to CANDELS magnitudes during flux calibration (see Figure 7 Right).

extraction profile (see Section 3.3 and Figure 5). To do this scaling we first read in our slit-star 1D extracted spectra in counts s$^{-1}$ and cut the 20 pixels on either edge to avoid any extra noise in the spectrum and removed the edges where the filter transmission curve drops below 70%. We then multiplied our slit star by our response curve to place it in $f_\nu$ units (purple, Figure 7 Right, Top). We fit the continuum slope and shape of the slit star to the suite of Kurucz (1993) models to get a best estimate at the spectral type and converted the model to $f_\nu$ units. These best-fit models are reported in Table 3.

The *J*-band magnitudes of our slit stars are taken from the updated photometric catalog of Finkelstein et al. (2022) based on the HST/CANDELS photometric data (Table 3) and converted into $f_\nu$ units. We then read in each of the appropriate filter transmission curves and put them all in units of Å: the MOSFIRE *J*-band (McLean et al. 2012) and the HST $J_{F125W}$ transmission curves (Figure 7 Right, Bottom). We interpolate the MOSFIRE filter and the Kurucz model onto the slit-star wavelength array and integrate them both through the MOSFIRE filter to get the flux of each in $f_\nu$ units. We scale the Kurucz model with $\frac{f_{\rm slit\,star}}{f_{\rm model}}$. We then use the Hubble transmission curve wavelength array and interpolate both the slit star and the scaled Kurucz model onto this wavelength array. Since the Hubble *J*-band filter covers a wider wavelength range than the MOSFIRE one we need to make sure we are not inaccurately accounting for the flux difference and we need to extend our slit-star spectrum to cover the whole HST wavelength array. To do this we appended the scaled Kurucz model spectrum to the edges of our slit-star array to fill the full wavelength space of the HST filter (blue, Figure 7 Right, Top). We then integrated this extended slit star through the HST filter and compared this flux to that found for our slit star in the CANDELS photometry, providing a scale for our slit star of $\frac{f_{\rm CANDELS}}{f_{\rm extended\,slit\,star}}$ (orange, Figure 7 Top, Right). The measured ratios of the known *J*-band magnitudes to our calibrated fluxes of the slit continuum sources are up to ∼75% off from the CANDELS measurements. Scale factors that were implemented to correct for this are shown in Table 3. We note that these scale factors are high, but this includes correction for slit loss, observing conditions, ground versus space-based flux differences, and the ∼2× over-extraction of our flux with our triple-Gaussian profile. We also note that performing a traditional single-Gaussian extraction

Table 4
Standard Stars Used for Flux Calibration

| Star Name | Masks Observed | R.A. (J2000) | Decl. (J2000) | Magnitude (2MASS) | Type | Best Fit Kurucz Model |
|---|---|---|---|---|---|---|
| HIP98400 | E1, E2, G1 | 299.8962917 | 11.8893833 | 8.791 | A0V | kp00_9500[g40] |
| HIP68767 | E3, E4, E5, E6 | 211.1265833 | 21.3881389 | 8.138 | B8V | kp00_12000[g40] |
| HIP56736 | C1, C2, C3, C4, U1, U2, U3, U4 | 174.4908750 | 15.7768889 | 8.790 | A0V | kp00_9500[g40] |
| HIP91315 | G2 | 279.3895833 | 62.5265833 | 5.725 | A0V | kp00_9500[g40] |

**Note.** A list of standard stars observed for flux calibration purposes and the corresponding masks that they were used to calibrate. Standard star magnitudes and spectral types were obtained from the 2MASS classifications (Skrutskie et al. 2006). We have included the Kurucz (1993) model we used to mask out absorption features and calibrate our spectra shown in Figure 7 (left). Note that standard stars were observed during each night of observation when they were at an airmass similar to our science observations, and were selected based on the suggested list from the MOSFIRE Observer instructions.

profile reduced the amount we need to scale our flux calibration, as expected. Differences between results from our triple-Gaussian method and this single-Gaussian profile test on our final results are discussed in Section 5.2. Our final response curve is our original response curve scaled with this final scaling per mask to match the HST photometry which we then convert to $f_\lambda$ units and use to flux calibrate all of our data.

### 3.5. Stacking of Multiple Nights of Observation

Every spectrum on each night of observation was flux-calibrated individually using the above steps. As some of the science masks were observed for multiple nights we then combined those observations to increase the exposure time on those objects. Since as discussed in Section 3.3.1 we did not know precisely where our objects were in their slits as there was no discernible pattern to the offsets of objects in our masks, we had seven different centered extractions for each galaxy for each night. In order to stack the spectra of our galaxy, we must then perform every possible combination of these different extraction centers for all the nights observed. To combine these nights, we weighted them by the seeing calculated by the best-fit Gaussian peak fluxes of the slit continuum sources. When we observed a galaxy over three nights, with seven possible locations in the slit, we have a total of 343 possible combinations of the spectra. For objects we observed over two nights we had 49 different combined spectra, and seven spectra for those observed only once. Total exposure times for each of our sources are reported in Table 2, as well as average seeing for each object. Seeing values for individual masks are reported in Table 1. A discussion about how we determined the best combination is in the following section.

## 4. Methods

### 4.1. Line Detections from an Automated Line Search

For our 11 objects of interest, the stacking process described in Section 3 created 1393 total spectra. Visually identifying faint lines in this large data set is infeasible owing to the large number of spectra and the inadequacy of human eyes at distinguishing faint lines from noise peaks. Rather than relying on uncertain and arbitrary visual inspection of 2D spectra to identify plausible emission lines, we utilize an automated line-finding code first published in Larson et al. (2018) and outlined here. This code uses a Markov Chain Monte Carlo (MCMC) routine to fit a model that consists of a Gaussian line plus a continuum constant to a given wavelength range, with four free parameters: the continuum level, line central wavelength, line FWHM, and integrated line flux. While Ly$\alpha$ likely has an asymmetric profile, at our expected low S/N we do not expect to resolve this asymmetry (e.g., Finkelstein et al. 2013), so a Gaussian function is an appropriate fit to this data. To run the MCMC, we use an IDL implementation of the affine-invariant sampler (Goodman & Weare 2010) to sample the *posterior* similar to that used in Finkelstein et al. (2019), which is similar to the emcee package (Foreman-Mackey et al. 2013).

As we do not want to bias ourselves by only looking at a specific section of the spectrum for our emission-line features based on our photometric redshifts, we search the entire spectrum with our automated code. At each wavelength pixel we perform an initial S/N check (flux/error), and if this is greater than unity, we run our fitting routine (this "pre-check" is not required but helps with efficiency, as no believable emission feature would have S/N < 1 at the line center). We also impose priors on each of the free parameters designed to increase efficiency without affecting the posterior distributions. For the continuum constant, we let it vary between $\pm 1 \times 10^{-18}$ erg s$^{-1}$ cm$^{-2}$ Å$^{-1}$, values that are much larger than the $1\sigma$ noise level and are much higher than the typical continuum values for our high-redshift sources. We restrict the peak wavelength to be the wavelength at that pixel $\pm 1.3$ Å (one pixel resolution element), such that we fit a Gaussian within each pixel. We limit the FWHM to between 2.6 Å (66 km s$^{-1}$), which is twice the pixel scale (smaller than the instrumental resolution at $R \sim 3300$ and expected FWHM for the MOSFIRE instrument in the $J$ band; McLean et al. 2012), and the FWHM that would correspond to 2000 km s$^{-1}$ ($\sim$80 Å) as calculated by FWHM$_{\mathrm{max}}$ = 2000 km s$^{-1}$ $\frac{\lambda_{\mathrm{peak}}}{c}$ (where $c$ is the speed of light); this is usually larger than anything observed at low to high$z$ in star-forming galaxies. The line flux prior requires the flux to be greater than $10^{-20}$ erg s$^{-1}$ cm$^{-2}$ and less than $10^{-15}$ erg s$^{-1}$ cm$^{-2}$, well outside the expected range of line fluxes for our high-redshift sample. These parameters are listed in Table 5.

We first used the IDL routine mpfit, a Levenberg–Marquardt least-squares fitting routine that fits a Gaussian within the above parameters (Markwardt 2009). The results of this code are then used as the starting point for our MCMC code, which we run with 100,000 iterations and 100 walkers on each pixel for each of our 1393 spectra, and we determine real/best line fit results as outlined in the following section. For our final fit results we use the median value of the last 10,000 steps of our MCMC chain. To measure the error on our parameters, we use the robust_sigma calculation: using the median absolute deviation as the initial estimate, then weighting points using Tukey's biweight (Equation (9) from Beers et al. 1990).

**Table 5**
MCMC Fitting Parameters

| Line Property | Allowed Parameter Range |
|---|---|
| Continuum | $-1 \times 10^{-18} < C < 1 \times 10^{-18}$ (erg s$^{-1}$ cm$^{-2}$ Å$^{-1}$) |
| Wavelength | $\lambda_{\text{pixel}} \pm 1.3$ (Å) |
| FWHM | 2.6 (Å) < FWHM < 78 (Å) |
| Line flux | $10^{-20} <$ flux $< 10^{-15}$ (erg s$^{-1}$ cm$^{-2}$) |

**Note.** The starting fit parameters we use for the MCMC chain that fits a constant + Gaussian function at each wavelength pixel. Continuum limits are far above and below the average flux/pixel to limit the restrictions placed on our fits. Peak wavelength is restricted to be between the current pixel and one on either side as we step down the spectrum in wavelength space and perform the fits at each position. FWHM limits are based on instrumental resolution (McLean et al. 2012) and physical expectations for broad Lyα emission lines (66 km s$^{-1}$ < FWHM < 2000 km s$^{-1}$). Line flux limits are below our detection threshold and above peak flux of our spectra to reduce restrictions on the MCMC chain.

We place several constraints on the Gaussian fits, including the wavelength, comparison to neighboring pixels, and comparison to all combinations at that pixel. First, we mask out the edges of the wavelength range for the MOSFIRE *J* band, where the transmission curve falls below 70%. We then require that any detections be found in a blank part of the spectrum, one devoid of sky emission lines as determined in the process detailed in Section 4.2.1. A line is considered a detection when it is the highest-S/N detection of those of the 2 pixels on either side; this ensures that we are finding the correct center of our emission lines that have a width of multiple pixels. In order to determine when we have the correct combination of our multiple nights of observation, we compare detections at the same pixel across all possible combined spectra of that source and select the one with the highest-measured S/N. This indicates that we have correctly lined up our extraction profiles for each night with our targets in the slit, recovering the full flux of the emission line for each night (see Figure 5, right panel). All final detections found by the automated line-finding routine, ones that meet all the above selection criteria, are then evaluated by the criteria for significant emission lines detailed in the follow sections.

### 4.2. Criteria for Detection

We performed several tests to determine whether the emission lines detected by the automated line-finding routine were real or spurious. First, we included a measure of the sky and error at the location of the line to help rule out faint skyline detections (Section 4.2.1). Second, we performed a negative test to get a contamination rate and threshold (Section 4.2.2). Third, we used two different measurements of S/N (Sections 4.2.3 and 4.2.4). The three methods and the resulting criteria for real detections that we implemented are described below.

#### 4.2.1. Integrated rms

We first examined the wavelengths of the detected features and found that many putative "lines" were located on or near telluric emission-line residuals. As the noise is significantly increased on these sky features (or near to them, due to their broad Lorentzian wings), we decided to exclude specific wavelengths from our search that corresponded to these telluric emission features. We first created a median stack of the rms error spectrum for all galaxies in the mask. This was necessary, as the location of the slit in the mask determines the wavelength coverage, and not every slit was centrally located in the mask and thus did not cover the full *J*-band wavelength array.

To further identify the lines, we removed the interline background (rms level between the atmospheric lines) by fitting it with two linear functions, on either side of 1.28 μm, where visual examination showed the slopes to be different.

We then subtracted these fits and made another 0.25σ cut (a by-eye determined value that included most faint skylines), and we marked any pixels below this as true interline wavelengths, in which we believe real weak emission lines can be found, with the remaining pixels denoted as skyline wavelengths. Any lines with central wavelengths in the skyline regions were removed from further consideration. This process excludes 46% of our spectrum from our consideration, which could prevent us from finding possible emission lines on the edges of skylines, but this removed any potential contamination by spurious detections in our final sample. While this worked to reduce the inclusion of skyline-noise spurious sources in our results, it is imperfect, as often the broad wings of skylines were still being included in emission-line detections. To account for this, for each MCMC-discovered line we calculated the integrated rms value over the same pixels our MCMC routine fit (peak ± Gaussian σ) for the emission line (Figure 8, *y*-axis). We further consider these values below.

#### 4.2.2. Negative Test

To further quantify the amount of contamination by spurious sources in our final list of detections, we performed a "negative spectrum test," where we multiplied all of our 1D spectra by −1 and then reran our full analysis, including stacking the negative spectra together across multiple nights and running our automated fitting routine in the same way as with our real data. Spurious detections due to noise should be symmetric about zero, while true detections will always be positive; this test helps diagnose the likelihood of a spurious signal being picked up by our detection routine. As shown in Figure 8 and discussed below, even after masking out skyline wavelengths, these negative "detections" occur predominantly in regions of the spectrum with high integrated rms values, further highlighting the utility of this quantity. We make a cutoff for our detection criteria (horizontal line in Figure 8) at the lowest rms of the detected negative lines, as these are not distinguishable from contamination.

#### 4.2.3. Integrated S/N

We used multiple ways of measuring the S/N of our emission-line detections to determine the robustness of our results. The first of these is the integrated S/N that comes from the Gaussian fit of the emission line from our automated line-finding code. We calculate the S/N of the emission line as the median line flux divided by the line flux error. This is the same way S/N was measured using this automated code in our Larson et al. (2018) paper, and a line passes this cut if it was found at >5σ in the stacked 1D spectrum (see vertical line in Figure 8). We also required that a real emission line be found in at least two individual nights of observation (all of our targets that only have one night of observation are devoid of any emission-line detections) at >3σ. Lines found in only one night at >3σ are indicated by circles in Figure 8, and lines detected in at least two individual nights are shown as stars.

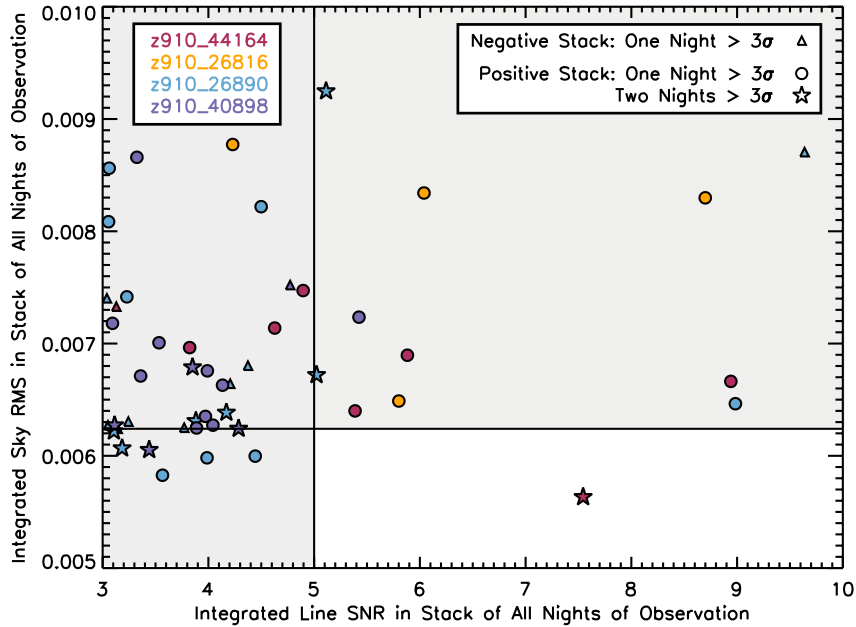

**Figure 8.** Integrated rms vs. S/N for detections in our automated emission-line search for all four galaxies in our final sample that had detections: z910_44164, z910_26816, z910_26890, z910_40898, all of which were observed at least twice. We require S/N > 5 for a significant detection (vertical line). We plot those detections found in at least one night at >3σ as circles and those found in at least two nights as stars. We ran our same automated line-finding routine on a negative spectrum (data × −1) and plot any detections found given the same criteria as triangles. We make our cut for a real detection at a point below the lowest detection in our negative spectra (horizontal line). Only one emission line found by our automated search meets all our criteria. Details of these thresholds can be found in Section 4.2, and this detection can be found in Figure 9 and Section 5.2.

### 4.2.4. Peak S/N

We did note that some sources with integrated S/N > 5 looked spurious upon visual inspection; thus, we also explore the S/N of the peak of the line, calculated by taking the maximum flux in the emission line (restricted to the center of the emission-line profile found by our code ±2 pixels) and calculating the ratio of this to the noise per pixel within our fitting range. This includes 1 pixel on either side of the center and serves as a way to calculate the flux of our emission line above the noise in the data at that wavelength range, independent of the Gaussian fit. Many of our detections were likely noise spikes in the data, or serendipitous peaks made by stacking a noisy pixel in one night with a noisy pixel in another. This quantity was used as an initial cut, as any lines that had a peak S/N < 5 were excluded as potential detections. All detections shown in Figure 8 have peak S/N > 5 in the fit to the stacked 1D spectrum.

## 5. Results

### 5.1. Identifying Real Emission Features

While our automated MCMC line detection routine found many potential features, we only consider lines to be real candidate emission lines if they have a *low* integrated rms value and a *high* S/N (integrated and peak) value, where skylines would fall in the category of high integrated rms and/or low S/N. Figure 8 shows all of our MCMC-identified putative emission lines (after removing those masked out on sky features), plotting their integrated S/N versus their integrated rms. We note that potential emission lines were only detected in four of our targets: EGS_44164, EGS_26816, EGS_26890, and EGS_40898, all of which were observed at least twice. True astrophysical emission features should occupy the lower right region of this figure, having low rms and high S/N. To decide exactly where to draw a boundary in these two quantities, we utilize the negative-spectra detections, which are shown by the triangles. These tend to populate the lower integrated S/N portion of the plot, though some do have higher integrated S/N. However, they do all fall above an integrated rms value of 0.00625 counts s$^{-1}$.

We thus initially consider features with integrated S/N > 5 and integrated rms < 0.00625 counts as potentially real emission features. We then consider that a line that has a significant S/N (>5σ) in the stacked spectrum should also be significantly (>3σ) detectable in each individual night of observation. In the figure, we plot emission lines found with S/N > 3 in one night as circles and those found with S/N > 3 in at least two nights as stars. As shown in Figure 8, there is only one emission feature with integrated S/N > 5 and integrated rms < 0.00625 counts detected at S/N > 3 in two nights. This emission feature comes from the galaxy EGS_z910_44164, is found at 11749 Å, has an integrated S/N of 7.5, and is detected at S/N > 3 in *both* nights of observation. We conclude that this emission feature is highly likely to be real. For the remainder of our observed galaxies, while some show potential emission features, they all have reasons to doubt their validity; thus, we consider them to be nondetections in our analysis. For the remainder of this paper we focus on this one likely real emission feature.

### 5.2. Detection of Lyα Emission at z = 8.665

The galaxy EGS_z910_44164 was observed in masks E1, E2, and E3, but we only include E1 and E3 in our final data set,

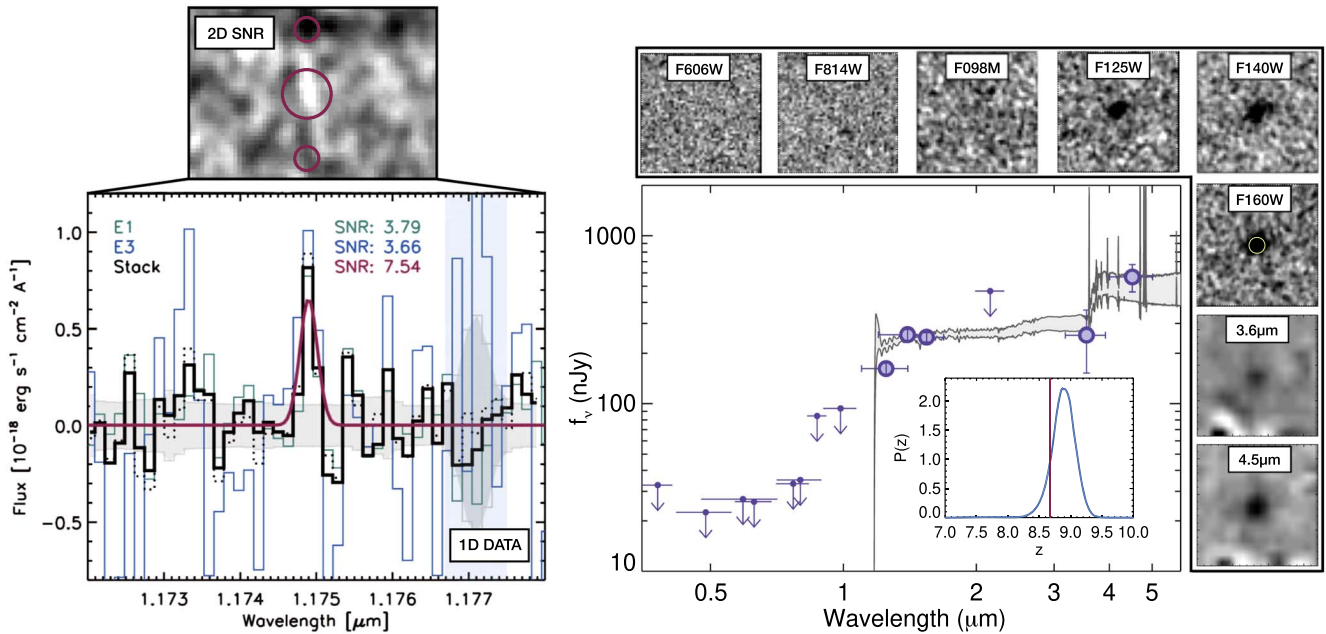

**Figure 9.** Left: the 1D stacked MOSFIRE *J*-band spectrum of EGS_z910_44164 (solid black with error in gray), with the automated line-finding code Gaussian fit (maroon) showing a significant ($7.54\sigma$) emission line detected at 11749.00 Å. The dotted black line shows the resulting stacked spectrum using a traditional single-Gaussian extraction profile as opposed to our triple-Gaussian profile as shown in Figure 5. This line is also detected at $>3\sigma$ in both nights of observation (blue and teal spectra), making it meet all our criteria for a robust detection. Details about this line can be found in Section 5.2. Stacked, smoothed 2D S/N spectra are shown above, where positive and negative traces of the emission line are highlighted by circles. Right: SED fit to the photometry for EGS_z910_44164 from HST and Spitzer (Tacchella et al. 2022) at the spectroscopic redshift measured with our data ($z = 8.665$). Postage stamp images for HST filters are $3'' \times 3''$. IRAC images are $10'' \times 10''$ and are neighbor subtracted.

as mask E2 was observed for less than 2 hr of integration time and was very noisy as a result of poor observing conditions (though we note that this line still meets all of our detection criteria for a real emission line when we include these data in our stacked spectrum). We show this object's 1D spectrum in the wavelength region around this line in Figure 9. This line is detected at $7.5\sigma$ in the stack (black) of our two nights of observation and at $3.8\sigma$ and $3.7\sigma$ in the individual nights (teal and blue). We measure the integrated flux of the emission line in the stacked spectrum to be $f_{\mathrm{Ly}\alpha} = (1.91 \pm 0.25) \times 10^{-18}$ erg s$^{-1}$ cm$^{-2}$ with an FWHM of $2.71 \pm 0.13$ Å. As a check to ensure that our triple-Gaussian extraction profile is not affecting our final flux measurement, we also performed a more traditional single-Gaussian extraction (and corresponding flux calibration) for this object and recover the same final line flux measurement. The stacked spectrum from this test is shown as a dashed black line in Figure 9. We note that this measured FWHM is below the instrumental resolution at this wavelength (3.5 Å), but this is not unexpected for low-S/N lines. Our uncertainty on the FWHM is artificially low, as our fitting routine prior restricts the FWHM to be larger than twice the pixel scale (2.6 Å) and the chain converges faster at the limit. We report the measured physical and emission-line properties for our detection of Ly$\alpha$ in EGS_z910_44164 in Table 6 and the photometric measurements for this galaxy in Table 7.

Interpreting this emission line as Ly$\alpha$, it would correspond to $z = 8.665 \pm 0.001$, and the rest-frame equivalent width (EW) would be $\mathrm{EW}_{\mathrm{Ly}\alpha} = 4.7 \pm 1.7$ Å, where we use the flux redward of Ly$\alpha$ in the spectrum of our best-fit spectral energy distribution (SED) model (Figure 9, right) as a measure of the continuum flux and $\mathrm{EW}_{\mathrm{Ly}\alpha} = \frac{f_{\mathrm{Ly}\alpha}}{f_{\mathrm{cont}}}$. To explore this

**Table 6**
The Measured Physical and Emission-line Properties for Our Detection of Ly$\alpha$ in EGS_z910_44164

| EGS_z910_44164 Properties | |
|---|---|
| R.A. (J200) | 215.218737 |
| Decl. (J200) | 53.069859 |
| $H_{160}$ | $25.41 \pm 0.07$ |
| $M_{\mathrm{UV}}$ | $-21.87 \pm 0.05$ |
| Emission-line Properties | |
| $z_{\mathrm{Ly}\alpha}$ | $8.665 \pm 0.001$ |
| $f_{\mathrm{Ly}\alpha}$ (erg s$^{-1}$ cm$^{-2}$) | $1.91 \pm 0.25 \times 10^{-18}$ |
| EW$_{\mathrm{Ly}\alpha}$ (Å) | $4.7 \pm 1.7$ |
| FWHM (Å) | $2.71 \pm 0.13$ (34.6 (km s$^{-1}$)) |
| Physical Properties | |
| log $M_\star$ ($M_\odot$) | $10.2^{+0.2}_{-0.2}$ |
| log SFR$_{50}$ ($M_\odot$ yr$^{-1}$) | $1.6^{+0.4}_{-0.3}$ |
| log sSFR$_{50}$ (Gyr$^{-1}$) | $0.5^{+0.4}_{-0.4}$ |
| $\rho_{\mathrm{UV}}$ (erg s$^{-1}$ Hz$^{-1}$) | $2.43 \pm 0.11 \times 10^{29}$ |

**Note.** Emission-line values obtained through a Gaussian fit to the emission-line feature in our MOSFIRE data. Physical properties for this galaxy were obtained through SED fitting with Prospector (Leja et al. 2017; Johnson et al. 2021) as described in Finkelstein et al. (2022) and Tacchella et al. (2022).

likelihood, we examined the photometric redshift probability distribution function $P(z)$ from Finkelstein et al. (2022), shown as the inset in Figure 9. This object exhibits a clear peak in its $P(z)$ at $z \sim 8.8$, with 99.5% of the integrated $P(z)$ lying at $z > 8$. The best-fit EAZY photometric redshift template at this redshift has $\chi^2 = 7.8$, compared to $\chi^2 = 31.3$ for the next-best redshift solution of $z \sim 1.7$. This strong preference for $z > 8$ is easily

**Table 7**
EGS_z910_44164 Photometric Measurements (in nJy)

| $V_{F606W}$ | $I_{F814W}$ | $Y_{F098M}$ | $J_{F125W}$ | $JH_{F140W}$ | $H_{F160W}$ | 3.6 $\mu$m | 4.5 $\mu$m |
|---|---|---|---|---|---|---|---|
| $-3.6 \pm 9.0$ | $1.2 \pm 11.7$ | $41.7 \pm 31.2$ | $161.9 \pm 14.0$ | $257.7 \pm 19.2$ | $249.2 \pm 15.5$ | $256.3 \pm 104.0$ | $568.0 \pm 105.5$ |

**Note.** Photometric values for EGS_z910_44164 in nJy for six HST filters from the CANDELS program (Grogin et al. 2011) and follow-up observations from Cycle 27 (PI Finkelstein). Spitzer/IRAC photometry is measured after deblending, and neighbor subtraction is performed with the TPHOT software (Merlin et al. 2016). Details on the photometric detection of this galaxy and how these values were obtained can be found in Finkelstein et al. (2022).

understandable upon examining this object's SED, also shown in the right panel of Figure 9. This object exhibits a clear Ly$\alpha$ break, being undetected at $\lambda \leqslant 1$ $\mu$m (including in follow-up F098M imaging), with strong detections redward of that point. In particular, for a low-redshift galaxy to mimic this Ly$\alpha$ break, it would have to be quite red, while the observed colors redward of the break are blue (e.g., $H - [3.6] = 0$). An alternative solution would be that this line is [O II] at $z = 2.15$, but this is a doublet and the other half would be located at either 11740 Å or 11759 Å, and we do not detect emission features in our spectrum at either of these wavelengths. We conclude that the photometry of this galaxy is inconsistent with any $z < 8$ and that our observed emission line is highly likely to be Ly$\alpha$.

Examining our sample, of the nine sources in the final Finkelstein et al. (2022) sample that we observed, this source has the brightest continuum flux. In fact, the only brighter source in the Finkelstein et al. (2022) final sample is their source EGS_z910_6811, which has Ly$\alpha$ confirmed by Zitrin et al. (2015). As a brighter source, we conclude that it is not surprising that EGS_z910_44164 is the only source in our observations from which we detected an emission line.

### 5.3. Properties of Detected Galaxy

We place this galaxy in context by comparing its physical properties to those of other very high redshift Ly$\alpha$-emitting galaxies. We use the results from Tacchella et al. (2022), who measured the stellar population properties of this galaxy fixing the redshift to the spectroscopic one measured from Ly$\alpha$ with the SED fitting code Prospector (Leja et al. 2017; Johnson et al. 2021). Using the results with their fiducial continuity prior on the star formation efficiency, we find that this galaxy has a stellar mass consistent with other $z > 7.5$ Ly$\alpha$ emitters in the literature, with a log $M_\star/M_\odot = 10.2^{+0.2}_{-0.2}$, a log SFR$_{50}$ of $1.6^{+0.4}_{-0.3}$ $M_\odot$ yr$^{-1}$, and a log sSFR$_{50}$ of $0.5^{+0.4}_{-0.4}$ Gyr$^{-1}$. There is another spectroscopically confirmed $z = 8.683$ galaxy in EGS from Zitrin et al. (2015) that we did not observe in this program, as Ly$\alpha$ was already detected at $f_{Ly\alpha} = 1.7 \times 10^{-17}$ erg s$^{-1}$ cm$^{-2}$ with an EW$_{Ly\alpha} = 28$ Å and an FWHM of 277 km s$^{-1}$. Zitrin et al. (2015) do not report physical properties of this galaxy, but Tacchella et al. (2022) fit the SED using the same technique and find a log $M_\star/M_\odot = 10.6^{+0.2}_{-0.3}$, slightly larger than that of EGS_z190_44164. We note that these results, especially the SFR and sSFR, can depend strongly on the star formation history priors. A detailed discussion on the impact different star formation histories have on the SED-measured properties of EGS_z190_44164 can be found in Figure 17 of Tacchella et al. (2022).

### 6. Discussion

#### 6.1. Likelihood of an Ionized Bubble at z = 8.7

As we likely would not expect to see Ly$\alpha$ at this redshift owing to the high neutral fraction of the IGM unless it occupied a large ionized bubble, we use the properties of this galaxy to determine whether it could have produced a large enough ionized region on its own for the Ly$\alpha$ to resonantly scatter out and become detectable. For this to happen, the ionized region needs to be ~1 physical (p)Mpc (Miralda-Escudé & Rees 1998; Malhotra & Rhoads 2002, 2006; Dijkstra 2014) in radius such that Ly$\alpha$ resonantly scatters out of resonance with the neutral IGM owing to the Hubble flow. We also note that this assumes no Ly$\alpha$ velocity offset from the intrinsic redshift of the galaxy, which would reduce the bubble size required for Ly$\alpha$ to transmit through the IGM, as a $\Delta v \gtrsim 300$ km s$^{-1}$ would scatter Ly$\alpha$ enough to escape from a <1 pMpc bubble (Mason & Gronke 2020). Alternatively, a measurement of the velocity offset compared to a systemic redshift could also inform on the Ly$\alpha$ transmission fraction through the IGM for a given bubble size, providing a better constraint on the intrinsic EW$_{Ly\alpha}$ as shown by Endsley et al. (2022). Future higher-S/N measurements of the line could probe the line shape, which would allow further constraints on the size of the bubble this galaxy inhabited. Currently, as we have no other measure of the systemic redshift of this galaxy, we show a conservative transmission radius for Ly$\alpha$ of 1 pMpc in Figure 10.

We use a similar method to that of Endsley et al. (2021) to calculate the expected radius of an ionized bubble that would be produced by our target source. Endsley et al. (2021) assume an ionizing photon escape fraction ($f_{esc}$) of 20% for their analysis, but as we do not have any measure for this value, we plot our bubble radii in Figure 10 as a function of escape fraction. From the observed brightness of our galaxy ($m_{F160W} = 25.41$ and $M_{UV} = -21.87$) and its spectroscopic redshift, we calculate the specific nonionizing UV luminosity of this galaxy to be $\rho_{UV} = (2.43 \pm 0.11) \times 10^{29}$ erg s$^{-1}$ Hz$^{-1}$ Mpc$^{-3}$.

To convert from this quantity to the intrinsic ionizing emissivity, we must first assume an ionizing photon production efficiency ($\xi_{ion}$), with units of erg$^{-1}$ Hz. For our calculations we assumed $\xi_{ion} = 25.6$, as this is consistent with the observations of Tang et al. (2021) in $z \sim 2$ analogs with EW$_{[O III]}$ similar to $z > 7$ galaxies. This value is also similar to values that have been observed for bright, $z \sim 7 - 8$ galaxies (Stark et al. 2015; Stark 2016) and is consistent with the redshift- and luminosity-dependent value predicted by Finkelstein et al. (2019). While the $\xi_{ion}$ versus EW distribution in the galaxies studied in Maseda et al. (2020) would imply that low-EW$_{Ly\alpha}$ galaxies would have $\xi_{ion}$ below the 25.6 value we use, we note that at our $M_{UV} = -21.8$, compared to the highest-redshift values in the study, a $\xi_{ion}$ of ~25.5 is not inconsistent with the observations included at $z \sim 5$. Values in this 25–26 range were also measured by Endsley et al. (2021) for their galaxies at $z \sim 6.8$, where they performed the same ionizing bubble radius calculation. Lacking a measurement of $\xi_{ion}$ for our galaxies, we assume one consistent with the literature but also discuss in Section 6.2 how our bubble size calculations would change if this value were higher than we estimated. The product of $\rho_{UV} \times \xi_{ion}$ gives this intrinsic ionizing emissivity ($\dot{N}_{ion}$, in units of s$^{-1}$). With this value we calculate $\dot{N}_{ion} = 10^{54.986}$ s$^{-1}$ for EGS_z910_44164. We note that the implied value used by Endsley et al. (2021) is $10^{55.08}$ s$^{-1}$ for their galaxies at $z \sim 6.8$.

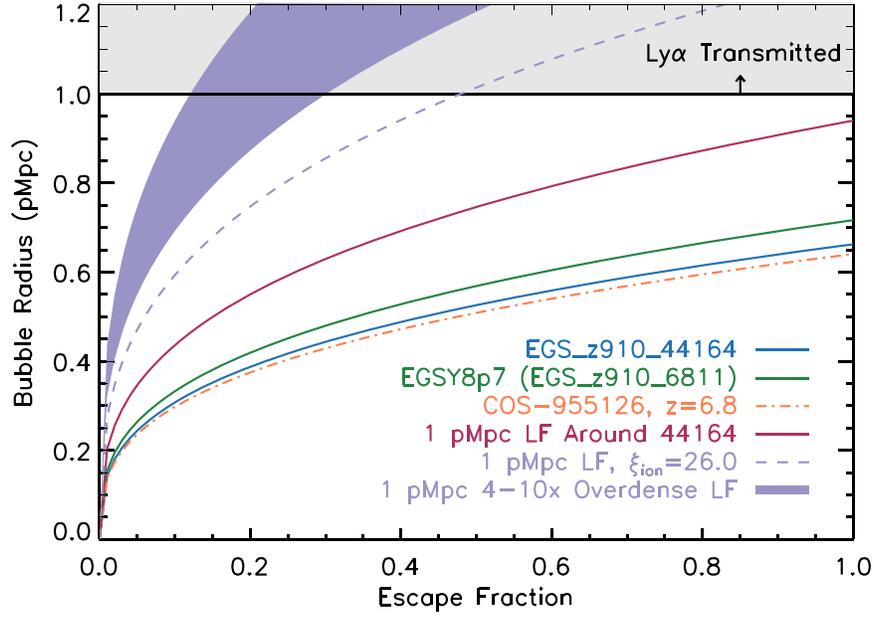

**Figure 10.** The calculated size of an ionized bubble produced by our galaxy, EGS_z910_44164 (blue), as a function of the ionizing escape fraction ($f_{esc}$). Bubble radii are calculated using Equation (4), where we assume a value of $\xi_{ion} = 25.6$ for all galaxies. For us to observe Ly$\alpha$ emission from a galaxy at this redshift, it must be located in an ionized region large enough that the Ly$\alpha$ photons resonantly scatter out of phase with the surrounding IGM, a region $\gtrsim 1$ pMpc as shown by the gray shaded portion. For our $z = 8.665$ galaxy, EGS_z910_44164, even with an escape fraction of 100% this galaxy alone does not produce a large enough ionized bubble. The same situation is true for the galaxy with Ly$\alpha$ detected by Zitrin et al. (2015), EGSY8p7 (green; EGS_z910_6811 in Finkelstein et al. 2022). Endsley et al. (2021) perform a similar calculation around their brightest source, COS-955126 at $z = 6.8$ (orange dashed). To account for the faint galaxies at this redshift that we are not detecting, we integrate the LF at $z = 8.665$ down to $M_{UV} = -13$ within 1 pMpc and add their contribution around EGS_z901_44164 (solid maroon). Our photometric redshift selection and the two spectroscopically confirmed targets indicate a 4–10× overdense region at the location of our objects, so we multiply the contribution the faint galaxies from the LF calculation by this factor, and the combined ionizing output can produce a 1 pMpc bubble with only a $f_{esc} \sim 12\%$–31% (purple shaded region). If we do not assume an overdensity but rather a higher $\xi_{ion} = 26.0$ value, then we can also produce a large enough bubble with $f_{esc} \sim 50\%$ (purple dashed). Both of these scenarios are possible but could only be confirmed with data from future telescopes such as JWST.

Following Cen & Haiman (2000) and Endsley et al. (2021), we calculate the size of an ionized bubble powered by this galaxy's ionizing emission as

$$R = \left( \frac{3 \dot{N}_{ion} f_{esc} t}{4\pi n_{HI}(z)} \right)^{1/3}, \quad (1)$$

where the proper volume density of neutral hydrogen is given by

$$n_{HI} = \frac{(1 - Y_{He}) \rho_{crit} \Omega_b}{m_p} (1 + z)^3 \quad (2)$$

with a helium mass fraction $Y_{He} = 0.2453$ (Planck Collaboration 2020) and critical density

$$\rho_{crit} = \frac{3 H_0^2}{8\pi G}. \quad (3)$$

The variable $t$ in Equation (1) represents the length of the current star formation episode in our galaxy. While Prospector does fit the star formation history, it is quite unconstrained, due to both the early epoch and the relatively small number of photometric constraints (Tacchella et al. 2022). We thus assume a star formation episode time of $t = 20$ Myr, but note that a longer star formation time would increase the size of our estimated bubbles. Endsley et al. (2021) assumed a 10 Myr timescale for their sample, with sSFR $\sim 5$ Gyr$^{-1}$; their values were input to show the bubble size of their galaxy (Figure 10, orange dashed line) in comparison to that which we calculate for our targets.

Putting all this together, and in terms of the values we used for our calculation for EGS_z910_44164, we can measure the bubble radius ($R$) in terms of our other unknowns, $\dot{N}_{ion}$, $f_{esc}$, and $t$:

$$R = 0.3875 \left[ \frac{\left(\frac{\dot{N}_{ion}}{10^{54.986} \text{ s}^{-1}}\right)\left(\frac{f_{esc}}{0.2}\right)\left(\frac{t}{20 \text{ Myr}}\right)}{\left(\frac{H_0}{67.36 \text{ km s}^{-1} \text{ Mpc}^{-1}}\right)^2 \left(\frac{\Omega_b}{0.0493}\right)\left(\frac{1+z}{9.665}\right)^3} \right]^{(1/3)} \text{ pMpc}.$$

$$(4)$$

For the ionizing output of the galaxy EGS_z910_44164 alone, this analysis gives a bubble size of 0.39 pMpc for $f_{esc} = 0.2$, or 0.66 pMpc for $f_{esc} = 1$ (thin blue line in Figure 10). Thus, even with a unity escape fraction, this bubble size is likely too small for Ly$\alpha$ to be transmitted since a $\sim 1$ Mpc bubble is needed for Ly$\alpha$ to resonantly scatter out of the IGM and be detectable (Malhotra & Rhoads 2004; Dijkstra 2014). However, this galaxy is not alone: 3.6 pMpc away from EGS_z910_44164 lies EGSY8p7 from Zitrin et al. (2015) at $z = 8.683$ (also known as EGS_z910_6811 from Finkelstein et al. 2022). We perform the same bubble size calculation for this source, accounting for its moderately higher UV luminosity ($M_{UV} = -22.13$; Finkelstein et al. 2022). As shown by the green line in Figure 10, for our assumed values the presumed bubble around EGSY8p7 is only slightly larger than that around EGS_z910_44164, reaching 0.72 pMpc for $f_{esc} = 1$ (or 0.42 pMpc for $f_{esc} = 0.2$). While it is possible that the ionizing photon production efficiency for EGSY8p7 may be larger owing to active galactic nucleus (AGN) activity (as indicated by a tentative N V detection in this galaxy by Mainali et al. 2018), it seems unlikely that either of these two bright galaxies alone can create large enough ionized bubbles to transmit Ly$\alpha$. Assuming a longer timescale for the star formation episode of, for example, 60 Myr could bring the bubble radius to the desired value. However, this would still require a unity escape fraction, which is unlikely. If this source does have an AGN, it could have a higher $\xi_{ion}$ value than we have used in this calculation owing to a harder ionizing spectrum, which could increase the ionized bubble size this galaxy could produce. The AGN could also increase the $f_{esc}$ for the galaxy, making a <1 pMpc bubble sufficient for Ly$\alpha$ transmission. Mainali et al. (2018) measure an Ly$\alpha$ velocity offset (from N V) of 362 km s$^{-1}$, which could also reduce the ionized bubble radius required for transmission through the IGM (Mason et al. 2018). In Figure 10 we show the bubble size around EGS_z910_6811 (green) assuming the same value for $\xi_{ion}$ (25.6) as with EGS_z910_44164, but all of the previously mentioned factors could increase the bubble radius, reduce the minimum bubble size needed for transmission, or make a higher escape fraction more likely. We do not include in this calculation the possibility that these galaxies are in close enough proximity that they would both contribute to the same ionized bubble, as the 1 pMpc bubbles centered on these two galaxies would not overlap at a distance of 3.6 pMpc.

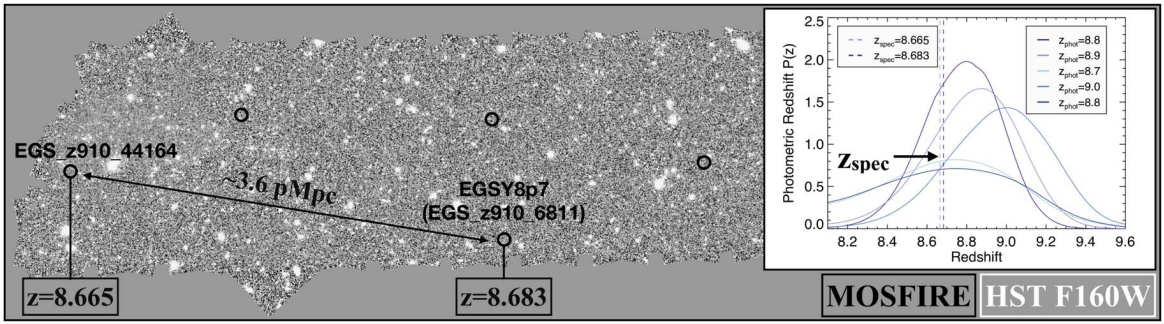

**Figure 11.** Overlay of the MOSFIRE pointings on the CANDELS EGS field. Our work with HST and Spitzer data indicates the presence of an overdensity at $z = 8.7$ in this field, via the discovery of five bright $z \sim 9$ galaxies (a factor of 4–10 more than expected at this redshift in a field this size; Figure 12). One of these galaxies was spectroscopically confirmed at $z = 8.683$ by Zitrin et al. (2015) (bottom middle); another is the galaxy we found to have Ly$\alpha$ at $z = 8.665$ (left). These two galaxies are physically separated by 3.6 pMpc, and their spectroscopic redshifts fall squarely in their photometric redshift distributions. $P(z)$ values for all five of these galaxies (EGS_z910_44164, EGS_z910_6811, EGS_z910_26890, EGS_z910_40898, EGS_z910_20381) place them at this same redshift (right), strong evidence for at least a 2$\sigma$ above-average density of galaxies at this redshift in a field this size.

While these two galaxies are bright, the local ionizing budget could be dominated by emission from fainter galaxies associated with these two objects. For example, Endsley et al. (2021) found that similarly bright galaxies at $z \sim 7$ can only ionize an $R = 0.3$ pMpc volume, similar to what we find here. The brightest of the Endsley et al. (2021) sources is plotted in Figure 10 as an orange dashed line, using their assumed values and cosmology. However, within a given volume around each galaxy there are fainter galaxies, which likely also contribute to (if not dominate) the ionizing emissivity. To account for the presence of these faint galaxies around EGS_z910_44164, we integrated the LF at $z = 8.665$ down to $M_{UV} = -13$ (assuming the Schechter form at $z = 8.665$ from Finkelstein 2016: $M^* = -20.44$, $\alpha = -2.20$, $\phi^* = 1.33 \times 10^{-3}$) and a volume with radius of 1 pMpc. In a random volume of the universe of this size, this LF predicts the presence of only 0.001 galaxies as bright as EGS_z910_44164. Therefore, to properly estimate the total UV luminosity density in a volume with radius 1 pMpc around this galaxy, we calculated the luminosity produced by all the galaxies in our LF and add in the remaining $0.999 \times$ our measured specific UV luminosity density for EGS_z910_44164. This gives a full-population $\rho_{UV}$ 2.87× higher than just EGS_z910_44164 can produce alone, producing a bubble with a radius of 0.55 pMpc with $f_{esc} = 0.2$, or 0.94 pMpc at $f_{esc} = 1.0$, still assuming $\xi_{ion} = 25.6$ erg$^{-1}$ Hz for all galaxies (maroon solid line in Figure 10). This bubble is still smaller than the $\sim 1$ pMpc sized bubble needed for us to detect the Ly$\alpha$ emission we see from these galaxies, even with a very high average escape fraction.

### 6.2. Implications of an Overdensity

This analysis suggests that even when accounting for the nearby fainter galaxies around our bright source, it is unlikely that an ionized bubble large enough could have been formed, even with extremely high ionizing photon escape fractions if this is a region of average galaxy density for this redshift. However, observations do suggest that this galaxy lives in a special place owing to the nearby presence of EGSY8p7 (EGS_z910_6811) and the fact that 7 of the 11 candidate bright $z = 9-10$ galaxies found by Finkelstein et al. (2022) are in the EGS field, with the remaining 4 across the other three CANDELS fields combined. Of the seven sources in the EGS field, the photometric redshifts for five of them (EGS_z910_44164, EGS_z910_6811, EGS_z910_26890, EGS_z910_40898, EGS_z910_20381) are consistent with the two spectroscopically confirmed galaxies with $8.6 < z_{phot} < 9.0$ as shown in Figure 11. While we observed four of these five galaxies (as EGS_z910_6811 already had a spectroscopic redshift from Zitrin et al. 2015), we only detect Ly$\alpha$ in the second-brightest galaxy, EGS_z910_44164. We also note that our observations of EGS_z910_26890 and EGS_z910_40898 had seeing on average over 1″, likely preventing detection of Ly$\alpha$, and our exposure time on EGS_z910_20381 was only ~3 hr, not long enough to detect Ly$\alpha$ even if it was at the same strength as the brightest galaxies in the sample.

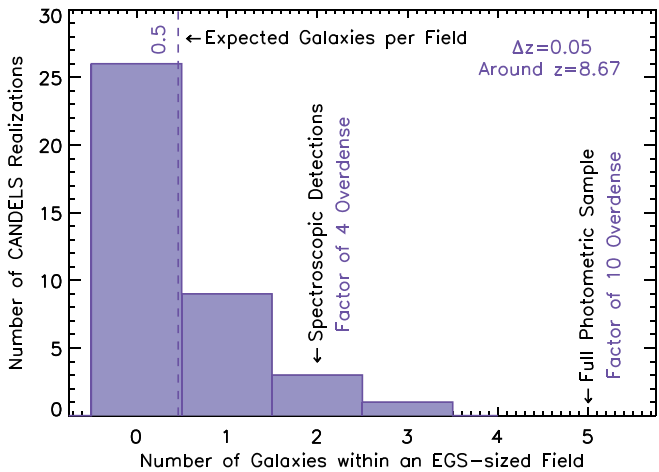

**Figure 12.** The expected number of galaxies per CANDELS-sized field, drawing samples from the Santa Cruz SAM (Yung et al. 2019). Our two sources with $z_s = 8.7$ are within $\Delta z = 0.05$, which is four times the expected number of galaxies in this sized field in this redshift range. This is a $2\sigma$-significance excess, including Poisson and cosmic variance uncertainties. If the three additional target sources in EGS with $z_p = 8.7$ have similar redshifts, this region would be a factor of 10 more overdense than the expected number in a $\Delta z = 0.05$ bin.

The presence of five bright galaxies all with photometric redshifts consistent with the two measured spectroscopic redshifts is highly suggestive that we are witnessing a potential candidate overdensity in the heart of the epoch of reionization. Similar to Finkelstein et al. (2022), we use a mock catalog based on the Santa Cruz semianalytic model (SAM; Yung et al. 2019) to explore the likelihood of this apparent overdensity representing a true physical structure, here focusing on $z \sim 8.7$. We excise 40 regions in this catalog, each the approximate size of a CANDELS field, and extract the number of $z \sim 8.7$ galaxies per field at $m_{F160W} \leqslant 26.5$ in two different redshift bins. We calculated the number of galaxies expected in a CANDELS-sized field in a $\Delta z = 0.05$ bin (Figure 12), the difference between our two spectroscopic redshifts centered around the median between them, $z = 8.67$, and find that the average expected number of galaxies at this spectroscopic redshift is 0.5 per CANDELS field to this depth. This is consistent with the number predicted by the evolving LF of Finkelstein (2016). However, with this technique, we can explore the significance of our overdensity inclusive of both Poisson and cosmic variance uncertainties. As shown in Figure 12, our spectroscopic detection of two galaxies in a single field is a $2\sigma$ outlier, with only 1/40 extracted regions encompassing more galaxies, making this field overdense by a factor of four. We also note that this overdensity factor is consistent with the 3–9x factor measured by Leonova et al. (2021) in this same field, using different methods. Even further evidence supporting the presence of an overdensity in the EGS field.

Our photometric observations of five galaxies in this field suggest that this field may be overdense by a factor of 10 if all of these galaxies lie in the same $\Delta z = 0.05$ bin as the two with spectroscopic redshifts. We also did this same count in a $\Delta z$ bin of 0.1, incorporating the potential spread in redshift space of the remaining photometric sources, and find an expectation value of one galaxy in a field this size. This would make our region overdense by a factor of five if all five sources lie within the larger $\Delta z$ range of 0.1. Folding this into our bubble size calculations by multiplying the total luminosity density calculated using our LF by an overdensity factor of 4–10 (still assuming the same $\xi_{ion}$ value), we find that enough ionizing photons can be produced to make a 1 pMpc bubble with $f_{esc} = 0.12$–0.31 (shaded purple region in Figure 10).

Lacking this overdensity, a higher value of $\xi_{ion}$ for all galaxies could also make a large enough bubble. As shown by the dashed purple line in Figure 10, assuming $\xi_{ion} = 26.0$ for all galaxies would produce a 1 pMpc bubble with $f_{esc} \geqslant 0.49$. While some low-mass compact star-forming galaxies in the low-redshift universe have been observed with such high $\xi_{ion}$ values (e.g., Izotov et al. 2017), it remains to be seen whether this is common for the entire high-redshift population. We also note that $f_{esc}$ values of 0.2 are higher than the typical $f_{esc}$ measured for the average low-redshift population. More data on these galaxies' properties from future telescopes such as JWST are needed to determine the appropriate values for these galaxies.

We conclude that the detection of Ly$\alpha$ emission from two galaxies in close proximity, EGS_z910_44164 and EGSY8p7, could be explained by the presence of an ionized region associated with a significant galaxy overdensity powering an ionized bubble early in the epoch of reionization (Figure 10). We do note that we have assumed that a bubble of 1 pMpc must be present for Ly$\alpha$ to be detectable. However, the Ly$\alpha$ from this galaxy is fairly weak; thus, it is also possible that the intrinsic Ly$\alpha$ is much higher, and only a small fraction is escaping through a much smaller ionized region. Simulations of this overdensity and confirmation of the redshift of the remaining galaxies in the sample would produce a more robust picture of the ionization state of this region.

## 7. Conclusions

We have presented the results of a large, systematic, spectroscopic follow-up survey of bright galaxy candidates at $z = 9$–10 using Keck/MOSFIRE observations. We obtained spectra for 12 of the 14 target galaxies from Finkelstein et al. (2022) across four CANDELS fields with exposure times between 3 and 10.5 hr with seeing typically under 1″. We used a combination of the MOSFIRE data reduction pipeline and our own methods to account for and correct for the drift of our objects in our slits throughout a night of observation. Our improved data reduction method also incorporated an extraction profile for spectra from 2D to 1D that is optimized for MOSFIRE's dithering pattern. Our flux calibration process used spectra of stars in our observation masks as an added measure of precision over just using a standard star obtained on the same night. Due to an offset of our objects from their expected position, and such faint, distant targets that visual inspection cannot locate them, we used a rolling extraction center procedure to ensure that we captured the full flux of our galaxies.

Building off of our automated line-finding routine from Larson et al. (2018), we searched through all possible combinations of our nights of observations for significant emission-line features. An intensive vetting process that required that a line pass multiple criteria before being determined robust yielded a single emission-line detection in galaxy

EGS_z910_44164 at 11749 Å. Our measurements, combined with deep HST and Spitzer photometry, indicate that this emission line is Lyα at $z = 8.665$ with a line flux of $(1.91 \pm 0.25) \times 10^{-18}$ erg s$^{-1}$ cm$^{-2}$ and EW$_{Ly\alpha}$ = 4.7 ± 1.7 Å. This is the second-brightest galaxy in the Finkelstein et al. (2022) sample, and the second to yield a detectable Lyα emission line, with the first being the brightest object in their sample, EGS_z910_44164 (EGSY7p8), with an Lyα line detected by Zitrin et al. (2015) at $z = 8.683$.

We explored whether our galaxy could possibly ionize a bubble around itself large enough that this Lyα emission would resonantly scatter and transmit through the IGM (1 pMpc; Miralda-Escudé & Rees 1998; Malhotra & Rhoads 2002, 2006; Dijkstra 2014). Neither our galaxy nor the brighter nearby EGS_z910_6811 could solely ionize enough of the IGM around them such that we would detect their Lyα emission. We included the contribution of the faint galaxies expected to be in the vicinity by integrating the LF from Finkelstein (2016), but even these combined with our bright galaxy produced an insufficient amount of ionizing radiation. Both of these objects are located in the CANDELS EGS field, and they are only two of the five sources with photometric redshifts in the same range, indicating the potential for an overdensity at $z = 8.7$. Including a 4–10× overdensity factor in our bubble size calculation, it is possible to produce an island of ionization in the early universe that Lyα could pass through to reach our detectors while only needing an ionizing photon escape fraction of ∼12%–31%.

From this analysis we conclude that the detection of Lyα from our source is suggestive that this galaxy exists in an ionized bubble early in the epoch of reionization. Under our fairly simple assumptions, we find that such a bubble should only be possible if powered by a significant overdensity of galaxies. This is consistent with the idea of "inside-out" reionization, where the most overdense regions in the universe ionize first, with bubbles eventually growing and overlapping to ionizing larger-scale regions. While constraining the overall ionized fraction from this single bubble is not possible, should future wide-field observations (such as deep galaxy surveys with the Roman Space Telescope followed up spectroscopically with telescopes such as the Giant Magellan Telescope) find that these bubbles are fairly common, it would imply a higher ionized fraction at $z \sim 9$ than predicted by typical late-reionization models.

We do note that we have made a variety of assumptions in this discussion. First, our implication that the observation of Lyα implies a 1 pMpc ionized bubble depends on both the amount of intrinsic Lyα emitted by the galaxy and how the kinematics in the interstellar medium of this galaxy affected the emission wavelengths of Lyα. Should the kinematics impart a significant net redshift onto Lyα (as is often seen at lower redshifts; e.g., Shapley et al. 2003; Song et al. 2014; Park et al. 2021), the Lyα emission could shift out of resonance within a smaller ionized bubble. Additionally, it is possible that the intrinsic Lyα strength is much higher than we observe (due to, e.g., a high ISM covering fraction and/or significant dust attenuation—though measurements by Tacchella et al. 2022 imply only a modest amount of attenuation in this source), which could imply significant IGM absorption, leading again to a smaller bubble size. Furthermore, with the available photometry we do not currently have strong constraints on $\xi_{ion}$ or the length of the current star formation episode ($t$) at this redshift, so we have assumed values consistent with similarly bright galaxies at slightly lower redshifts, but we show the impact different assumptions would have on our results in Equation (4). Finally, we conclude that the escape of Lyα from these two galaxies could be due to the presence of an ionized bubble produced by a larger number of galaxies in an associated overdensity at $z = 8.7$.

All of these assumptions will soon be testable with observations from JWST. Spectroscopy with NIRSpec will allow modeling of both the ionizing environment within this galaxy and the kinematic state within the interstellar medium. This deep spectroscopy can also probe weaker Lyα emission from the remaining bright $z \sim 9$ candidates. Spectroscopy with MIRI will allow a reconstruction of the intrinsic Lyα emission via Balmer line measurements. Finally, deep NIRCam imaging and spectroscopic follow-up will confirm (or reject) the presence of this overdensity. The forthcoming data from the Cosmic Evolution Early Release Science (CEERS)[17] Survey (Finkelstein et al. 2017) will contain observations of this field—providing answers to some of our proposed questions with first light on JWST.

R.L.L., S.L.F., and M.B. acknowledge that they work at an institution, the University of Texas at Austin, that sits on indigenous land. The Tonkawa lived in central Texas, and the Comanche and Apache moved through this area. We pay our respects to all the American Indian and Indigenous Peoples and communities who have been or have become a part of these lands and territories in Texas. We are grateful to be able to live, work, collaborate, and learn on this piece of Turtle Island.

The authors wish to recognize and acknowledge the very significant cultural role and reverence that the summit of Maunakea has always had within the indigenous Hawaiian community. We are most fortunate to have the opportunity to conduct observations from this mountain.

This work was supported by NASA Keck PI Data Awards administered by the NASA Exoplanet Science Institute, PIDS 77/2018A_N132 (PI R. Larson), 68/2018B_N145 (PI R. Larson), 61/2019A_N079, 58/2019B_N117 (PI R. Larson), and 78/2020A_N063 (PI T. Hutchison). Data presented herein were obtained at the W. M. Keck Observatory from telescope time allocated to the National Aeronautics and Space Administration through the agency's scientific partnership with the California Institute of Technology and the University of California. The Observatory was made possible by the generous financial support of the W. M. Keck Foundation.

This material is based on work supported by the National Science Foundation Graduate Research Fellowship under grant Nos. DGE-1610403 and DGE-1746932.

R.L.L and S.L.F. acknowledge support from NASA through ADAP award 80NSSC18K0954. T.A.H. and C.J.P. acknowledge generous support from Texas A&M University and the George P. and Cynthia Woods Institute for Fundamental Physics and Astronomy.

This publication makes use of data products from the Two Micron All Sky Survey, which is a joint project of the University of Massachusetts and the Infrared Processing and Analysis Center/California Institute of Technology, funded by the National Aeronautics and Space Administration and the National Science Foundation.

---

[17] https://ceers.github.io/

Table 8
Observed Galaxies No Longer in Target Sample

| ID | R.A. (J2000) | Decl. (J2000) | $H_{F160W}$ (mag) | Masks Observed | Total Exp. Time (hr) | Avg. Seeing (arcsec) | Reason Removed from Sample |
|---|---|---|---|---|---|---|---|
| EGS_z910_1977 | 215.1845833 | 52.9817222 | 26.99 | E1, (E2) | 4.90 (6.23) | 0.64 (0.8) | $mag_{F160W}$ too faint |
| EGS_z910_31527 | 215.1170000 | 53.0274444 | 26.51 | E1, (E2) | 4.90 (6.23) | 0.64 (0.8) | S/N($J$, $H$) = 6.8 (too low) |
| EGS_z910_35139 | 215.1916250 | 53.0712861 | 25.52 | E1, (E2) | 4.90 (6.23) | 0.64 (0.8) | bad IRAC deblending |
| GOODSN_z910_52354 | 189.2177083 | 62.3117778 | 26.79 | (G1) | (0.99) | (0.70) | $mag_{F160W}$ too faint |
| GOODSN_z910_39040 | 189.2556250 | 62.3552111 | 25.35 | (G1) | (0.99) | (0.70) | S/N($J$, $H$) = 5.0 (too low) |
| GOODSN_z910_50839 | 189.2150417 | 62.3179389 | 26.62 | (G1) | (0.99) | (0.70) | $\Delta\chi^2 = 2.6$ |
| GOODSN_z910_50980 | 189.3795000 | 62.3174333 | 26.29 | (G1) | (0.99) | (0.70) | $\Delta\chi^2 = 0.0$ |
| GOODSN_z910_52539 | 189.2122917 | 62.3124000 | 26.73 | (G1) | (0.99) | (0.70) | $mag_{F160W}$ too faint |
| UDS_z910_34040 | 34.3291667 | −5.2008417 | 26.55 | U1, U2, U3, U4 | 18.23 | 0.92 | bad HST deblending |
| UDS_z910_26785 | 34.3561667 | −5.1856139 | 26.55 | U1, U2 | 10.41 | 1.04 | $\Delta\chi^2 = 1.7$ |
| UDS_z910_65630 | 34.3567500 | −5.1607861 | 26.39 | U1, U2 | 10.41 | 1.04 | $\Delta\chi^2 = 0.9$ |
| UDS_z910_67665 | 34.3333333 | −5.1642000 | 26.97 | U1, U2 | 10.41 | 1.04 | $mag_{F160W}$ too faint |
| COSMOS_z910_4158 | 150.0727500 | 2.1860833 | 26.05 | C1, C2 | 5.37 | 0.77 | $\Delta\chi^2 = 1.6$ |
| COSMOS_z910_5039 | 150.0720833 | 2.1903250 | 25.00 | C2 | 3.25 | 0.83 | high persistence |
| COSMOS_z910_52242 | 150.0662500 | 2.4086306 | 26.27 | C3, (C4) | 2.98 (3.25) | 0.64 (0.76) | $\Delta\chi^2 = -5.63$ |

**Note.** Galaxies observed during this Keck program but that are no longer in the sample after further cuts were made to the sample selection for our $z = 9-10$ targets. All IDs are from the Finkelstein et al. (2022) catalog. Cuts required $mag_{F160W} < 26.6$, S/N($J$, $H$) > 7, and $\Delta\chi^2$ between the high- and low-redshift solutions >3.5, as well as successful deblending in both HST and Spitzer images and a check for persistence left from prior observations of bright sources at the same pixel location as the target. During our data reduction and analysis, we did not include observations from masks with less than 2 hr of exposure time, as we found the data to be too noisy and the scientific value negligible. These masks, exposure times, and seeing are included for completeness but indicated by parentheticals.

The authors acknowledge the Texas Advanced Computing Center (TACC) at The University of Texas at Austin for providing database, and grid resources that have contributed to the research results reported within this paper http://www.tacc.utexas.edu.

The authors would like to thank all of our Keck Observatory Support Astronomers, Josh Walawender, Sherry Yeh, Alessandro Rettura, and Elena Manjavacas, for their assistance during our observing runs. Special thanks to all of our Operating Assistants for their work driving the telescope during our observations: John Pellitier, Alan Hatakeyama, Joel Aycock, Julie Renaud-Kim, Heather Hershley, Terry Stickel, and Arina Rostopchina.

*Facilities:* HST, Spitzer, Keck Observatory, Texas Advanced Computing Center.

*Software:* TPHOT (Merlin et al. 2016), mpfit (Markwardt 2009), LA Cosmic, (van Dokkum 2001), DRP (http://keck-datareductionpipelines.github.io/MosfireDRP/), Prospector (Leja et al. 2017; Johnson et al. 2021).

## Appendix
## Sources Observed That Are No Longer in the $z = 9-10$ Sample

During the time period of our observations, we were still finalizing our selection process of $z = 9-10$ sources in the CANDELS fields. Thus, we targeted 15 sources in our MOSFIRE masks that we later removed from our final photometric sample. Only those remaining after all cuts were made and additional space- and ground-based photometry were acquired are included in the main text of this paper. Here we list those sources that were not in the final sample of Finkelstein et al. (2022), including the reason they were removed, but which we did observe. It should also be noted that we reduced the data for these galaxies and performed the same search for emission lines as for those still in our sample, but we found no significant emission features. The selection criteria for our sample are detailed in Finkelstein et al. (2022) and described briefly in the main text. These galaxies were removed because they either did not have required $mag_{F160W} < 26.6$, S/N($J$, $H$) > 7, or have $\Delta\chi^2$ between the high- and low-redshift solutions >3.5, did not have successful deblending in the HST or Spitzer images, or were contaminated in the imaging by persistence left from prior observations of bright sources at the same pixel location as the target. The location of our G1 masks was chosen, as it included a high number of candidate sources in our sample at the time; unfortunately, all of those sources were later removed for the reasons listed in Table 8.

### ORCID iDs

Rebecca L. Larson https://orcid.org/0000-0003-2366-8858
Steven L. Finkelstein https://orcid.org/0000-0001-8519-1130
Taylor A. Hutchison https://orcid.org/0000-0001-6251-4988
Casey Papovich https://orcid.org/0000-0001-7503-8482
Micaela Bagley https://orcid.org/0000-0002-9921-9218
Mark Dickinson https://orcid.org/0000-0001-5414-5131
Sofía Rojas-Ruiz https://orcid.org/0000-0003-2349-9310
Harry C. Ferguson https://orcid.org/0000-0001-7113-2738
Intae Jung https://orcid.org/0000-0003-1187-4240
Mauro Giavalisco https://orcid.org/0000-0002-7831-8751
Andrea Grazian https://orcid.org/0000-0002-5688-0663
Laura Pentericci https://orcid.org/0000-0001-8940-6768
Sandro Tacchella https://orcid.org/0000-0002-8224-4505